\documentclass[aps,prb,preprint,groupedaddress,floatfix]{revtex4}
\usepackage{graphicx,amssymb,color,array}
\begin{document}
\title{The central role of entropy in adiabatic ensembles and its application to phase transitions in the grand-isobaric adiabatic ensemble.}
\author{Caroline Desgranges and Jerome Delhommelle}
\affiliation{Department of Chemistry, New York University, New York, New York 10003, United States}
\affiliation{Department of Chemistry, University of North Dakota, Grand Forks ND 58202, United States}

\date{\today}

\begin{abstract}
Entropy has become increasingly central to characterize, understand and even guide assembly, self-organization and phase transition processes. In this work, we build on the analogous role of partition functions (or free energies) in isothermal ensembles and that of entropy in adiabatic ensembles. In particular, we show that the grand-isobaric adiabatic $(\mu,P,R)$ ensemble, or Ray ensemble, provides a direct route to determine the entropy. This allows us to follow the variations of entropy with the thermodynamic conditions and thus to explore phase transitions. We test this approach by carrying out Monte Carlo simulations on Argon and Copper in bulk phases and at phase boundaries and assess the reliability and accuracy of the method through comparisons with the results from flat-histogram simulations in isothermal ensembles and with the experimental data. Advantages of the approach are multifold and include the direct determination of the $\mu-P$ relation, without any evaluation of pressure via the virial expression, the precise control of the system size and of the number of atoms via the input value of $R$, and the straightforward computation of enthalpy differences for isentropic processes, which are key quantities to determine the efficiency of thermodynamic cycles. A new insight brought by these simulations is the  highly symmetric pattern exhibited by both systems along the transition, as shown by scaled temperature-entropy and pressure-entropy plots.
\end{abstract}

\maketitle

\section{Introduction}
Using statistical mechanics and thermodynamics, eight ensembles can be defined~\cite{fowlerstatistical,hill1986introduction,graben1993eight,escobedo2006simulation}. The first four to be introduced~\cite{McQuarrie} were the microcanonical ensemble $(N,V,E)$, canonical ensemble $(N,V,T)$, grand-canonical ensemble $(\mu,V,T)$ and the isothermal-isobaric $(N,P,T)$ ensemble. These ensembles have been used extensively in computer simulations~\cite{Allen} to model a wide range of continuous or discrete systems and are at the basis of molecular dynamics~\cite{nose1984molecular,andersen1980molecular,ccaugin1991molecular,martyna1994constant,martyna1996explicit,bond1999nose,bussi2007canonical,tuckerman2006liouville} or Monte Carlo methods~\cite{mcdonald1972npt,jorgensen1982convergence,ray1991microcanonical,ray1996microcanonical,mezei1987grand,orkoulas1999phase,smit1995grand,jorgensen1998temperature,errington1998phase}. A fifth ensemble, known as the generalized $(\mu,P,T)$ ensemble, was later proposed  by Guggenheim~\cite{guggenheim1939grand}. The canonical, grand-canonical isothermal-isobaric and generalized ensembles form a set of four isothermal ensembles, for which $T$ is held constant. However, by modifying the thermodynamic constraints, it is possible to define three additional statistical ensembles. Using the Legendre-Laplace mapping procedure, Brown, Hill and Ray developed three other ensembles, known as the isoenthalpic-isobaric ensemble $(N,P,H)$~\cite{brown1958constant,haile1980isoenthalpic}, the grand-isochoric adiabatic $(\mu,V,L)$ ensemble~\cite{ray1981new,kristof1996alternative,graben1991unified} and the grand-isobaric adiabatic $(\mu,P,R)$ ensemble~\cite{ray1990fourth,graben1991unified,ray1993monte}. Together with the microcanonical ensemble, they form a set of four adiabatic ensembles, in which the value taken by a heat function, rather than of the temperature, is constant. In short, the isothermal ensembles are in thermal contact with a reservoir, while the adiabatic ensembles are thermally insulated.

A striking feature of adiabatic ensembles is the direct connection they provide with entropy. In isothermal ensembles, the partition function is related to the free energy. For instance, the Helmholtz free energy $A$ can be written as $-\beta A(N,V,T)=\ln Q(N,V,T)$, in which Q(N,V,T) is the canonical partition function. Similarly, the Gibbs free energy $G$ is given by $-\beta G(N,P,T)=\ln Q(N,P,T)$ with Q(N,P,T) the isothermal-isobaric partition function. We also have for the Landau free energy, or grand potential $(J)$, $-\beta J(\mu,V,T)=\ln Q(\mu,V,T)$ in the grand-canonical ensemble and for the null potential $(Z)$, $-\beta Z(\mu,P,T= \ln Q(\mu,P,T)$ in the Guggenheim ensemble. However, in adiabatic ensembles, the partition function becomes directly related to the entropy. For instance, in the microcanonical $(N,V,E)$ ensemble, the entropy is equal to $S(N,V,E)=k_B \ln Q(N,V,E)$. This is the famous Boltzmann relation for the entropy~\cite{McQuarrie}. A similar relation actually holds for all adiabatic ensembles~\cite{fowlerstatistical,hill1986introduction,graben1993eight,escobedo2006simulation}. In the isoenthalpic-isobaric ensemble $(N,P,H)$, we have $S(N,P,H)= k_B \ln Q(N,P,H)$, while in the ($\mu$,V,L) and ($\mu$,P,R) ensembles, we find $S(\mu,V,L)= k_B \ln Q(\mu,V,L)$ and $S(\mu,P,R)= k_B \ln Q(\mu,P,R)$, respectively. This direct connection with entropy makes adiabatic ensembles especially promising for the study of phase transitions, in which entropy is a key property to detect and follow the order $\leftrightarrow$ disorder transitions. 

In particular, the grand-isobaric adiabatic $(\mu,P,R)$ ensemble gives a direct access to the value taken by the entropy and, as a result, is especially well suited to study such transitions. However, there have been very few calculations performed in the $(\mu,P,R)$ ensemble~\cite{ray1993monte,ray1993new} and none so far, to our knowledge, for a system undergoing a phase transition. The goal of this work is thus to show how simulations in the ($\mu$,P,R) adiabatic ensemble enable the determination of the properties of bulk phases, on the example of the liquid and of the vapor phase, and of the conditions for which the vapor-liquid phase transition takes place. Specifically, for each value of the heat function $R$, we can calculate the entropy as the ratio $S={R \over T}$, in which $T$ is evaluated through the equipartition principle, and determine $S$ for a wide range of thermodynamic conditions such as, {\it e.g.}, along isobars and to detect the onset of first-order phase transitions (through the observation of discontinuities in the entropy), and of second-order phase transitions, and the presence of a critical point (through the continuous variation of $S$). This provides a complete picture of the phase transition process from an entropic viewpoint.

The paper is organized as follows. We start by introducing, in the next section, the formalism underlying adiabatic ensembles, starting from the well-known microcanonical ensemble, and most particularly, the grand-isobaric adiabatic $(\mu,P,R)$ ensemble. We also discuss how the entropy is evaluated in this adiabatic ensemble and how $(\mu,P,R)$ simulations can be implemented within a Monte Carlo framework. Section III provides an account of the results obtained for bulk phases, for the vapor-liquid phase transition and for the onset of supercritical behavior through the determination of the entropy of the system on the examples of Argon, modeled with the Lennard-Jones potential, and of Copper, modeled with a many-body, embedded atoms model. We highlight the role played by the heat function and how fine-tuning the choice for the input value of $R$ leads to a precise control of the system size. We also assess the validity and reliability of $(\mu,P,R)$ simulations in the case of Argon by comparing the simulation results to the experimental data and to the results obtained in prior work using flat-histogram simulations, while Copper provides an interesting application of the approach since the determination of the critical properties of metals is a notoriously difficult task. Furthermore, we discuss how such simulations can be used to evaluate enthalpy difference during isentropic processes, which, in turn, allow for the direct determination of the efficiency of thermodynamic cycles. We finally draw the main conclusions from this work in the last section.

\section{Formalism and simulation methods}

\subsection{From the microcanonical ensemble to the grand-isobaric adiabatic ensemble}

We provide in this section a brief account of the principles and formalism underlying adiabatic ensembles and, in particular, the $(\mu,P,R)$ ensemble. We start by introducing the formalism in the case of the most well-known adiabatic ensemble, {\it i.e.} the microcanonical ensemble. Adiabatic ensembles emerge from the combination of the zeroth, first and second laws of thermodynamics~\cite{graben1991unified}. This means that thermodynamic equilibrium between the system and its surroundings is characterized by three macroscopic variables characterizing the onset of thermal, mechanical and chemical equilibrium, respectively. Unlike isothermal ensembles, in which temperature serves as the thermal equilibrium variable, adiabatic ensembles use a heat function as a thermal variable. In the microcanonical $(N,V,E)$ ensemble, the system is completely isolated from its surroundings and there is no heat exchange or matter exchange. As a result, the entropy $S$ only depends on $N$,$V$ and $E$ through
\begin{equation}
dS=(1/T)dE + (P/T)dV - (\mu/T)dN
\label{micro}
\end{equation}
Equilibrium is reached when $dE=0$, $dV=0$, $dN=0$ and, hence, $dS(N,V,E)=0$. This equation defines $S(N,V,E)$, in which $N$ is the chemical equilibrium variable, $V$ the mechanical equilibrium variable and the heat function $E$ (or internal energy) is the thermal equilibrium variable.

Let us now consider the grand-isobaric adiabatic $(\mu,P,R)$ ensemble. In this ensemble, fluctuations in the number of particles $N$ and in the volume $V$ are allowed. This means that that the system is adiabatically insulated from the reservoir and that $\mu$ and $P$ are kept constant through the action of a porous adiabatic piston. For this ensemble, the Ray energy $R$ plays the role of the heat function. $R$ is defined as
 \begin{equation}
R=E+PV-\mu N
\end{equation}
The entropy $S$ now depends on $\mu$,$P$ and $R$ through
\begin{equation}
dS=(1/T)dR - (V/T)dP - (N/T)d\mu
\label{Ray}
\end{equation}
Here, equilibrium is reached when $dP=0$, $d\mu=0$, $dR= dE + VdP -\mu dN=0$, leading to $dS(\mu,P,R)=0$.

\subsection{Entropy calculations in adiabatic ensembles}

Legendre transforms are often used in thermodynamics to link different state functions such as, for instance, the internal energy and the enthalpy. They are especially useful in providing a connection between the thermodynamic potentials for isothermal ensembles and for adiabatic ensembles. The Legendre transform of the entropy $S(N,V,E)$ yields the following equation between the thermodynamic potential for the canonical ensemble, {\it i.e.} the Helmholtz free energy $A(N,V,T)$, and that for the microcanonical ensemble, {\it i.e.} the entropy $S(N,V,E)$
\begin{equation}
-\beta A(N,V,T)=k_B^{-1} S(N,V,E) -\beta E
\label{Legendre1}
\end{equation}
in which $\beta=1/(k_BT)$, leading to the well-known relation $A=E-TS$.

We now turn to the grand-isobaric adiabatic $(\mu,P,R)$ ensemble and its isothermal counterpart, the generalized $(\mu,P,T)$ ensemble introduced by Guggenheim. Carrying out a Legendre transform of the entropy $S(\mu,P,R)$ provides a relation with the Guggenheim thermodynamic potential $Z(\mu,P,T)$ through 
\begin{equation}
-\beta Z(\mu,P,T)=k_B^{-1} S(\mu,P,R) -\beta R
\label{Legendre2}
\end{equation}
From a statistical standpoint, the Legendre transforms of Eqs.~\ref{Legendre1} and~\ref{Legendre2} also provide a connection between the partition functions of isothermal ensembles and the phase space volume, or number of microstates, of adiabatic ensembles. For instance, Eq.~\ref{Legendre1} relates $A(N,V,T)=-k_BT \ln Q(N,V,T)$ and $S=k_B \ln Q(N,V,E)$, where the latter is the well-known Boltzmann formula and $Q(N,V,E)$ denotes the number of microstates. Similarly, in the $(\mu,P,R)$ ensemble, specifying $R$ will characterize a macrostate, that consists of a very large number of microstates. The computation of the number of such microstates can then yield the entropy of the system through $S(\mu,P,R)=k_B \ln Q(\mu,P,R)$. A key advantage of the $(\mu,P,R)$ ensemble is that it provides a much more straightforward way to evaluate the entropy. Indeed, as discussed in previous work\cite{graben1993eight}, $Z(\mu,P,T)=0$. This leads to the following relation
\begin{equation}
k_B^{-1} S(\mu,P,R) =\beta R
\end{equation}
This means that the entropy $S$ can be obtained from $R$ in the $(\mu,P,R)$ ensemble through
\begin{equation}
S(\mu,P,R)= {R \over T}
\label{entropy}
\end{equation}
This means that simulations in the $(\mu,P,R)$ ensemble can readily provide the value for the entropy of the system. In such simulations, $R$ is an input parameter, that remains constant throughout the simulation, and the average temperature can be computed over the course of the simulations using the equipartition principle, leading to a straightforward determination of $S$. In the next subsection, we discuss how we implement Monte Carlo (MC) simulations in the $(\mu,P,R)$ ensemble and determine the entropy of a system.

\subsection{Monte Carlo simulations in the $(\mu,P,R)$ ensemble}

In the $(\mu,P,R)$ ensemble, the chemical potential $\mu$, pressure $P$ and heat function $R$ are constant, which means that the temperature $T$, number of particles $N$ and volume $V$ are allowed to fluctuate. This means that there are four types of Monte Carlo (MC) moves for simulations in this ensemble: (i) random displacements of particles, (ii) deletion of a randomly chose particle, (iii) insertion of a new particle at a random position in the system and (iv) a random volume change for the system. To determine the acceptance rules for each of the MC moves, we start by defining the probability associated with a configuration containing $N$ particles with coordinates ${\mathbf q}$ in a volume $V$ as
\begin{equation}
P({\mathbf q}, N, V)= {(bV)^N \over \Gamma(3N/2) Q(\mu,P,R)} \left[ R-PV+\mu N-U({\mathbf q})\right]^{3N/2-1}
\end{equation}
where $b=(2\pi m / h^2)^{3/2}$. This means that, for a MC move corresponding to a random displacement of a particle, starting from an old $(o)$ set to a new $(n)$ set of coordinates $(\mathbf q_o) \to (\mathbf q_n)$, the Metropolis method yields the following acceptance rule
\begin{equation}
acc (o \to n) = min \left[ 1, {P({\mathbf q_{n}}, N, V) \over P({\mathbf q_o}, N, V)} \right] = min \left[ 1,  {\left[ R-PV+\mu N-U({\mathbf q_n})\right]^{3N/2-1} \over  \left[ R-PV+\mu N-U({\mathbf q_o})\right]^{3N/2-1}} \right]
\end{equation}

The acceptance rule for the deletion of a randomly chosen particle {\it i.e.} from $(\mathbf q_o,N) \to (\mathbf q_n,(N-1))$, follows as 
\begin{equation}
acc (o \to n) = min \left[ 1,   {{N \Gamma(3N/2)} \over {bV\Gamma(3(N-1)/2)}} \times {{[R-PV+\mu (N-1) -U({\mathbf q_n})]^{3(N-1)/2-1}} \over {[R-PV+\mu N-U({\mathbf q_o})]^{3N/2-1}}} \right]
\end{equation}

The acceptance rule for the insertion of a particle at a random position {\it i.e.} from $(\mathbf q_o,N) \to (\mathbf q_n,(N+1))$, is given by
\begin{equation}
acc (o \to n) = min \left[ 1,  {{bV \Gamma(3N/2)} \over {(N+1) \Gamma(3(N+1)/2)}} \times {{[R-PV+\mu (N+1) -U({\mathbf q_n})]^{3(N+1)/2-1}} \over {[R-PV+\mu N-U({\mathbf q_o})}]^{3N/2-1}}  \right]
\end{equation}

Finally, for a random volume change from $(\mathbf q_o,V_o) \to (\mathbf q_n,V_n)$, the Metropolis criterion becomes
\begin{equation}
acc (o \to n) = min \left[ 1,  { V_n^N \left[ R-PV_n+\mu N-U({\mathbf q_n})\right]^{3N/2-1} \over  V_o^N \left[ R-PV_o+\mu N-U({\mathbf q_o})\right]^{3N/2-1}} \right]
\end{equation}

\subsection{Simulation details}

Argon is modeled with the Lennard-Jones potential, as given by
\begin{equation}
\phi(r_{ij})=4\epsilon \left[ {\left({\sigma \over r_{ij}}\right) ^{12}- \left({\sigma \over r_{ij}}\right) ^6} \right]
\label{LJ}   
\end{equation}
where $r_{ij}$ is the distance between the two atoms, and $\epsilon$ and $\sigma$ denote the energy and size parameters for the potential. Here we use the following set of parameters, with $(\epsilon/k_B)=117.05$~K and $\sigma=3.4$~\AA, which performs very well~\cite{errington2003direct,PartI} against the experimental data for the vapor-liquid equilibrium properties of Argon~\cite{Vargaftik}. The calculation of the interactions between pairs of Ar atoms are truncated beyond a distance of $3\sigma$ and the conventional tail corrections are added to the overall energy of the system~\cite{Allen}.

We model $Cu$, and the many-body interactions that take place in this metal, with an embedded-atom potential (EAM)~\cite{finnis1984simple,sutton1990long,mei1991analytic,daw1983semiempirical} known as the  quantum-corrected Sutton-Chen~\cite{luo2003maximum} embedded atoms (qSC-EAM) potential. The qSC-EAM potential is a density-dependent force fields that has been shown to be very versatile, as it models accurately the thermodynamic and transport properties of liquid metals~\cite{luo2003maximum,desgranges2008rheology,kart2005thermodynamical,xu2005assessment,desgranges2014unraveling,desgranges2018unusual,desgranges2019can}, as well as their boiling points~\cite{gelb2011boiling,Tsvetan2}. According to this force field, the potential energy $U$ of a system containing $N$ atoms is equal to the sum of a two-body term and of a many-body term

\begin{equation}
U={1 \over 2} \sum_{i=1}^N \sum_{j \ne i}{\varepsilon \left( {a \over r_{ij}} \right)^n} -\varepsilon C \sum_{i=1}^N \sqrt \rho_i
\label{total}   
\end{equation}
in which $r_{ij}$ is the distance between two atoms $i$ and $j$ and the density term $\rho_i$ is given by
\begin{equation}
\rho_i=\sum_{j \ne i}{\left( {a \over r_{ij}} \right)^m}
\label{dens}   
\end{equation}

We use  the parameters obtained by Luo {\it et al.}~\cite{luo2003maximum} for  $Cu$, with $\varepsilon=0.57921\times10^{-2}$~eV, $C=84.843$, $a=3.603$~\AA, $n=10$ and $m=5$. The cutoff distance is set to twice the parameter $a$ as in previous work~\cite{luo2003maximum}. 

Simulations in the $(\mu,P,R)$ ensemble are carried out as follows. The MC moves are attempted according to the following rates: (i) $33$\% of attempted moves are random displacements of a particle, (ii) $33$\% are random insertions , (iii) $33$\% are deletions of randomly selected particles within the system, and (iv) $1$\% are random volume changes. We perform simulations over a wide range of conditions, that encompass the vapor-liquid coexistence and the supercritical domain of the phase diagram. We detail below in the results section how we choose specific values  for the input parameters, and examine specifically the role played by the heat function $R$ in the next section. For each set of conditions, we carry out two successive runs. We first perform a run of $10^8$ MC steps to allow the system to relax and the simulation to converge towards equilibrium, and then a run of $10^8$ MC steps over which averages are calculated. Statistical uncertainties are evaluated using the standard block averaging technique over blocks of $50 \times 10^7$ MC steps. In particular, the average temperature of the system is calculated by recognizing that, for each configuration, the kinetic energy $K$ is equal to
\begin{equation}
K = \left[ R-PV+\mu N- U\right]
\end{equation}
and by applying the equipartition principle through
\begin{equation}
<T>={ 2 <K> \over 3 <N> k_B}
\label{temp}
\end{equation}
This, in turn, allows for the determination of the entropy of the system through Eq.~\ref{entropy}.

\section{Results and Discussion}

We start by commenting on the results obtained for Argon, which provides an excellent test case to analyze the impact of the input parameters, and most particularly of the heat function $R$, on the fluid properties, and to assess the accuracy of simulations $(\mu,P,R)$ given the large amount of experimental and theoretical data available in the literature. To understand better the role of $R$, we begin by carrying out a series of simulations in the grand-isobaric adiabatic ensemble by varying the value specified for $R$ and keeping the values for $\mu$ and $P$ constant. This allows us to measure how a change in the input parameter $R$ (total value for the heat function or Ray energy) modifies the average value for the number of atoms $<N>$, for the temperature $<T>$, the specific volume $<V> \over <N>$, the enthalpy per atom $<H> \over <N>$, the Ray energy per atom $<R> \over <N>$ and the entropy per atom $<S> \over <N>$. We provide in Table~\ref{Tab1} the numerical results for a series of $(\mu,P,R)$ simulations with $\mu=-400$~kg/kJ, $P=1000$~bar and $R\over k_B$ ranging from $4 \times 10^5$~K to $2 \times 10^6$~K. Table~\ref{Tab1} shows that $<N>$ increases with $R$. For instance, when $R/k_B=4\times10^5$~K, we find $<N>=182.84 \pm 0.06$, while for $R/k_B=2\times10^6$~K, we obtain $<N>=913.83\pm 0.10$. This means that there is a 5-fold increase in $<N>$ when the input value for $R$ is multiplied by 5.  This allows for a direct control of the number of atoms in the system for $(\mu,P,R)$ simulations through the fine-tuning of the value of $R$ for a given set of $\mu$ and $P$ values. Table~\ref{Tab1} also shows that all intensive thermodynamic properties, including the temperature, specific volume, enthalpy per atom, Ray energy per atom and entropy per atom, converge towards the same value for a given pressure and chemical potential. Indeed, for $\mu=-400$~kg/kJ and $P=1000$~bar, we have $<T>= 224.4\pm0.2$~K, ${<V> \over <N>} = 0.877 \pm 0.001$~cm$^3$/g, ${<H> \over <N>} = 55.3 \pm 0.2$~kJ/kg, ${<R> \over <N>} = 455.05 \pm 0.15$~kJ/kg and ${<S> \over <N>} = 2.028 \pm 0.001$~kJ/kg/K obtained through ${R \over <N><T>}={S \over <N>}$. This is a key feature of $(\mu,P,R)$ simulations, since it allows to control the impact of finite-size effects on the simulation results and sample large enough system sizes, even when the density of the system is very low as, {\it e.g.}, for metallic vapors. This additional flexibility is particularly advantageous when compared to simulation methods, such as simulations in the grand-canonical ensemble, that keep the volume fixed and thus require a careful choice for $V$ when a wide density range needs to be sampled as, for instance, during a phase transition.

\begin{table}[h!]
\centering
\begin{tabular}{ccccccc}
\hline\hline
${R \over k_B}$~(K)~&~$<N>$~&~$<T>$~(K)~&~$<V> \over <N>$~(cm$^3$/g)~&~$<H> \over <N>$~(kJ/kg)~&~$R \over <N>$~(kJ/kg)~&~$<S> \over <N>$~(kJ/kg/K)\\
\hline
$4 \times 10^5$~& 182.84 & 224.27 & 0.877 & 55.11 & 454.94 & 2.029 \\
$5 \times 10^5$~& 228.41 & 224.57 & 0.877 & 55.49 & 455.22 & 2.027 \\
$6 \times 10^5$~& 274.08 & 224.54 & 0.877 & 55.08 & 455.23 & 2.027 \\
$7 \times 10^5$~& 319.86 & 224.35 & 0.877 & 55.27 & 455.09 & 2.029 \\
$8 \times 10^5$~& 365.47 & 224.46 & 0.877 & 55.46 & 455.19 & 2.028 \\
$9 \times 10^5$~& 411.18 & 224.29 & 0.877 & 55.19 & 455.16 & 2.029 \\
$1 \times 10^6$~& 456.94 & 224.33 & 0.877 & 55.09 & 455.10 & 2.029 \\
$2 \times 10^6$~& 913.83 & 224.29 & 0.877 & 55.19 & 455.12 & 2.029 \\
\hline\hline
\end{tabular}
\caption{Argon: $(\mu,P,R)$ simulation results for a chemical potential of $\mu=-400$~kg/kJ, a pressure of $P=1000$~bar and different values of the heat function $R$~(kJ/kg).}
\label{Tab1}
\end{table}

\begin{figure}
\begin{center}
\includegraphics*[width=10.8cm]{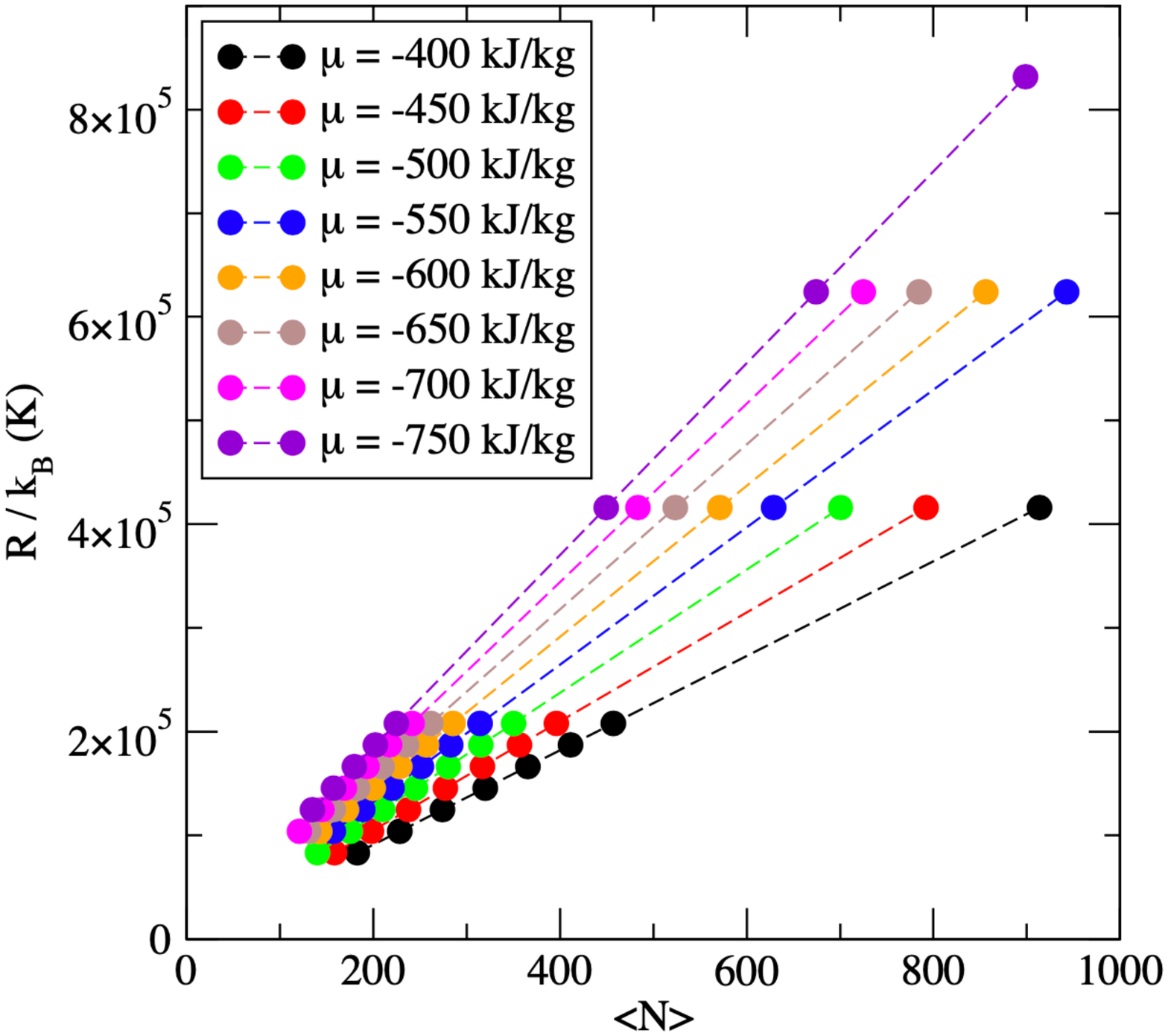}(a)
\includegraphics*[width=8.6cm]{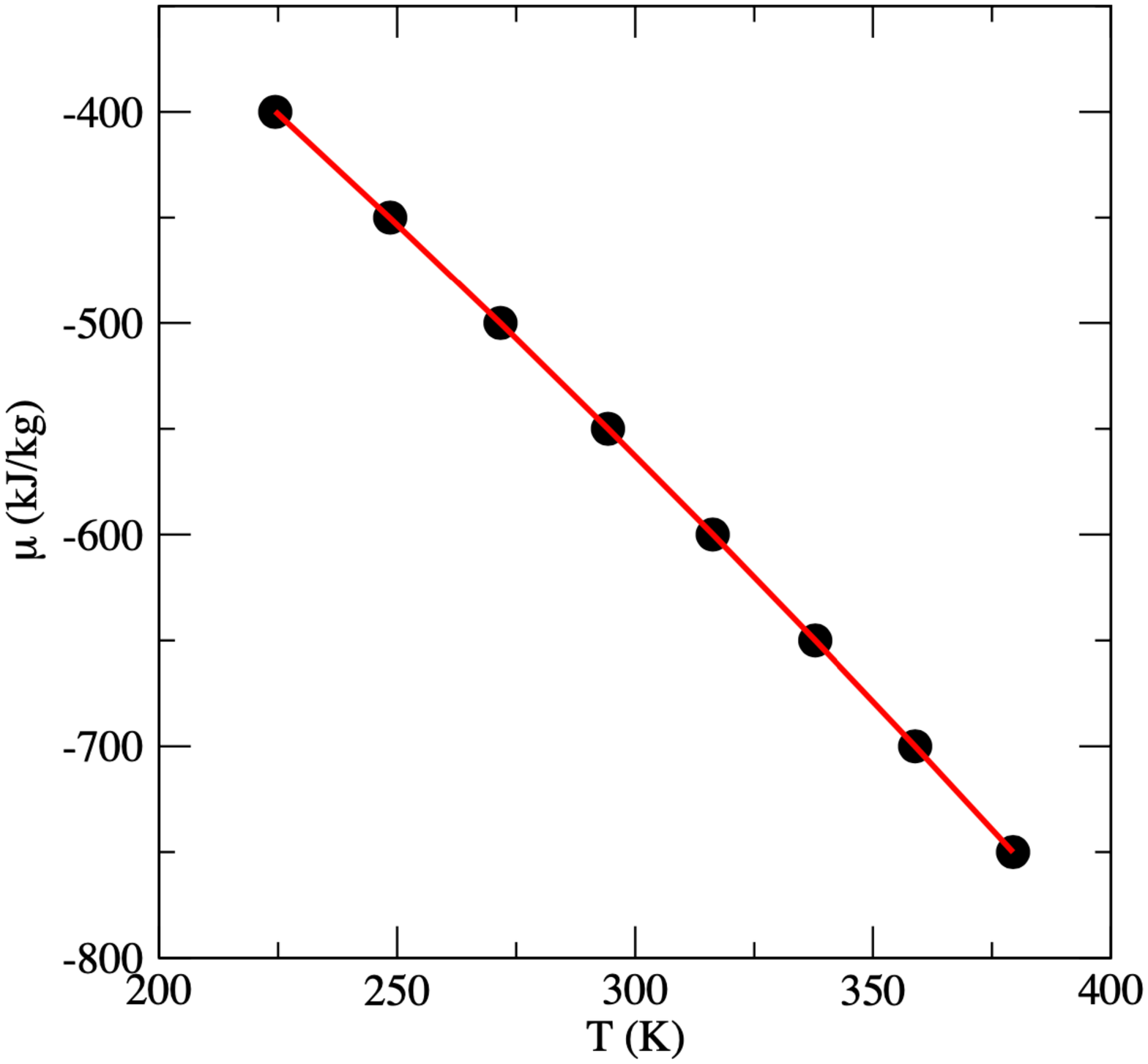}(b)
\end{center}
\caption{Argon along the $P=1000$~bar isobar: (a) Heat function or Ray energy ($R / \ k_B$) against the average number of atoms in the system ( $<N>$) for different values of $\mu$, and (b) $\mu$ against $<T>$, with a quadratic fit to the simulation results shown as a red solid line.}
\label{Fig1}
\end{figure}

To refine our understanding of the relation between $R$ and $S$ during $(\mu,P,R)$ simulations, we focus on a specific isobar, $P=1000$~bar, and modify the values for both $R$ and $\mu$. Fig~\ref{Fig1}(a) shows the results obtained for chemical potentials ranging from $\mu=-400$~kJ/kg to $\mu=-750$~kJ/kg. For each $\mu$, we vary the value of $R \over k_B$ over an energy interval ranging from $4 \times 10^5$~K to $2 \times 10^6~K$, and plot the results obtained for $R\over k_B$ as a function of the average number of atoms in the system $<N>$ in Fig.~\ref{Fig1}(a). This plot shows that the linear relation between $R$ and $<N>$ observed when $R$ varies and $\mu$ and $P$ are fixed holds for a wide range of $\mu$, with a slope $R \over <N>$ varying from $455.05$~kJ/kg to $925.48$ over the $\mu$ interval considered here, as reported in Table~\ref{Tab2}. Then, one only needs to calculate $<T>$ through Eq.~\ref{temp} over the course of the simulations to evaluate how entropy varies along the $P=1000$~bar isobar. We give in Table~\ref{Tab2} the values obtained for the entropy from the $(\mu,P,R)$ simulations. $<S> \over <N>$ is found to exhibit the expected trend, with a steady increase in entropy as the chemical potential decreases, {\it i.e.}, as the specific volume increases (or, equivalently as density decreases) and the enthalpy per atom increases.

\begin{table}[h!]
\centering
\begin{tabular}{ cccccc }
\hline\hline
$\mu$~(kJ/kg)~&~$<T>$~(K)~&~$<V> \over <N>$~(cm$^3$/g)~&~$<H> \over <N>$~(kJ/kg)~&~$R \over <N>$~(kJ/kg)~&~$<S> \over <N>$~(kJ/kg/K)\\
\hline
-400 & 224.4 & 0.877 & 55.3 & 455.05 & 2.028\\
-450 & 248.5 & 0.928 & 75.1 & 525.04 & 2.112\\
-500 & 271.7 & 0.979 & 93.8 & 593.76 & 2.185 \\
-550 & 294.3 & 1.030 & 111.6 & 661.63 & 2.248  \\
-600 & 316.3 & 1.080 & 128.6 & 728.59 & 2.304  \\
-650 & 337.8 & 1.131 & 145.0 & 794.92 & 2.353  \\
-700 & 358.9 & 1.181 & 160.6 & 860.53 & 2.398  \\
-750 & 379.4 & 1.230 & 175.5 & 925.48 & 2.439  \\
\hline\hline
\end{tabular}
\caption{Argon: entropy calculations via $(\mu,P,R)$ simulations along the $P=1000$~bar isobar.}
\label{Tab2}
\end{table}

Another way to calculate the entropy of a system consists in computing the dependence of the chemical potential upon temperature. We show in Fig~\ref{Fig1}(b) a plot of $\mu$ as a function of temperature when the pressure is held fixed at $P=1000$~bar. These results are obtained for similar system sizes, {\it i.e.} by setting a value for $R$ corresponding to a system size of about 400 atoms. Using the Gibbs-Duhem relation, we have $N d\mu = -S dT +V dP$, which, at constant pressure, can be written as $N d\mu = -S dT$ or $d\mu/dT = -S/N$. This implies that Fig~\ref{Fig1}(b) can provide a graphical estimate of the entropy. Performing a linear regression of the $\mu=f(T)$ plot provides a slope, and thus a rough estimate of $S$, of $2.26$~kJ/kg/K. This value is in good agreement with the results reported in Table~\ref{Tab2}, which show that $S$ varies from $2.028$~kJ/kg/K to $2.439$~kJ/kg/K over the range of chemical potential studied here. To provide a more accurate estimate for the entropy, we perform a quadratic fit of the plot, and find that, after differentiation with respect to $T$, the entropy follows the linear law: $S$~(kJ/kg/K~$=2.5932 \times 10^{-3} \times T + 1.4764$, which yields entropy values that are within $0.03$~kJ/kg/K of the $(\mu,P,R)$ simulation results reported in Table~\ref{Tab2}. This also provides a validation of the entropy values obtained for each individual $\mu$ value along the $P=1000$~bar isobar.

\begin{figure}
\begin{center}
\includegraphics*[width=10cm]{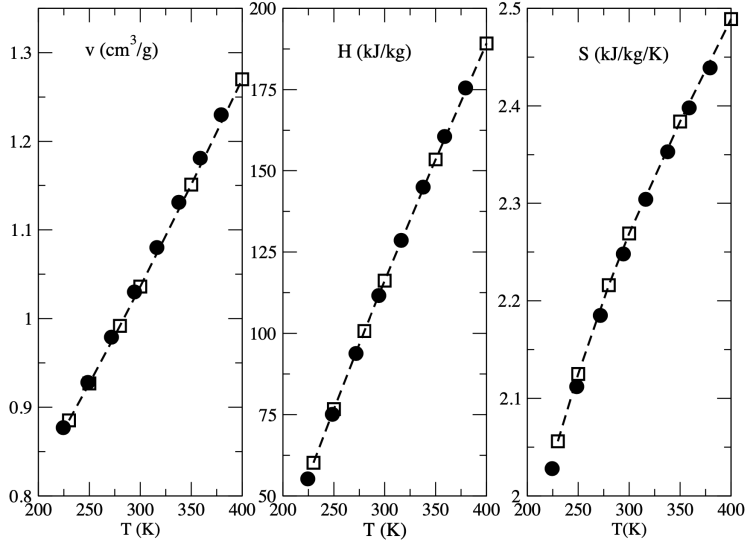}
\end{center}
\caption{Argon along the $P=1000$~bar isobar: dependence of several intensive properties, including the specific volume, enthalpy and entropy, upon temperature along the $P=1000$~bar isobar. Results from $(\mu,P,R)$ simulations are shown as filled circles and compared to the experimental data~\cite{Vargaftik}, shown as open squares.}
\label{Fig2}
\end{figure}

The next step is the assessment of the accuracy of the $(\mu,P,R)$ simulation results. For this purpose, we collect averages of several intensive properties, including the specific volume, enthalpy and entropy over the course of  $(\mu,P,R)$ simulation and compare the simulation results to the available experimental data~\cite{Vargaftik}. We provide a graphical account of this comparison in Fig.~\ref{Fig2}. As shown on this graph, the simulation results are in excellent agreement with the experimental data for Argon over the entire isobar. We observe the expected trend with an increase in the specific volume (or, equivalently, a decrease in the fluid density) with temperature. This leads to an increase in enthalpy with temperature, as a result of both the increase in kinetic energy with temperature and in the potential energy with the increase in specific volume, and thus with fewer and less attractive interatomic interactions taking place within the fluid. As for the entropy $S$, we observe a steady increase which is consistent with the increase in specific volume, and thus the greater number of microstates, or of possible atomic arrangements, for each of these macrostates. We add that the steady variations of the intensive properties along the $P=1000$~bar isobar are also consistent with the absence of any phase transitions. This is the case here, since for Argon~\cite{Vargaftik}, the critical parameters are $P_c=50$~bar and $T_c=150.86$~K, and thus, for the conditions studied in Fig.~\ref{Fig2}, Argon is a supercritical fluid.

\begin{figure}
\begin{center}
\includegraphics*[width=10cm]{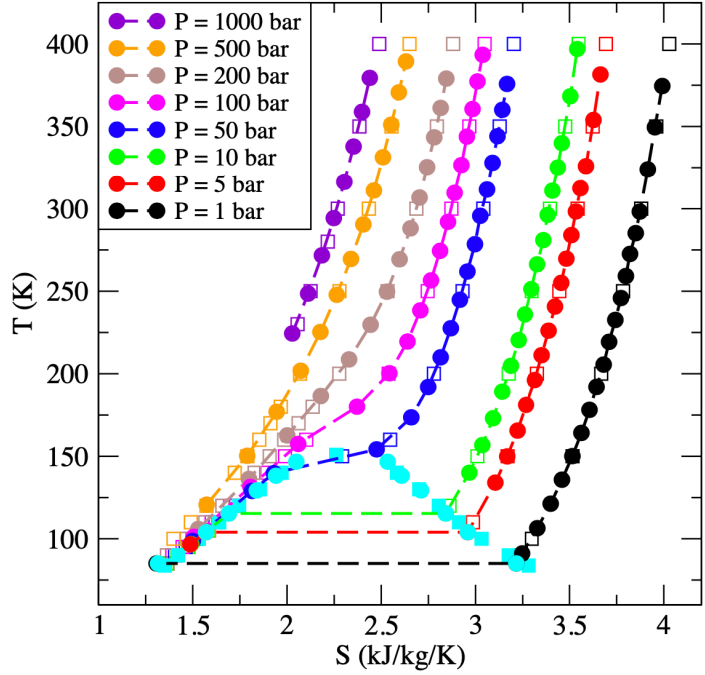}
\end{center}
\caption{Temperature-entropy plot for Argon. $(\mu,P,R)$ simulation results are shown as filled circles (colors vary according to the isobar they refer to, as indicated in the legend), while experimental data~\cite{Vargaftik} are shown as squares. Symbols in cyan outline the conditions for which vapor-liquid coexistence are found for the $(\mu,P,R)$ simulations (filled circles) and experimentally (filled squares). Only a few isobars are shown for clarity, while the $(\mu,P,R)$ results at coexistence are shown for all pressures considered here.}
\label{Fig3}
\end{figure}

We now examine how the $(\mu,P,R)$ simulation method performs for the detection and prediction of phase transitions and focus on the example of the vapor-liquid transition. We thus carry out $(\mu,P,R)$ simulations along isobars ranging from $P=1000$~bar to $P=1$~bar. For each isobar, we gradually increase the chemical potential from a low value, corresponding to a high temperature-low density fluid (see, for instance, the last row of Table~\ref{Tab2} with $\mu=-750$~kJ/kg for $P=1000$~bar isobar), to a high value of $\mu$, corresponding to a low temperature-high density fluid ({\it e.g.}, $\mu=-400$~kJ/kg for $P=1000$~bar isobar). As long as the input value for the pressure remains sufficiently high, the entropy of the system decreases steadily, as temperature decreases, along an isobar, and does not exhibit any break. This behavior can be seen on the temperature-entropy plot in Fig.~\ref{Fig3} for isobars such that $P>50$~bar. On the other hand, for isobars with $P<50$~bar, we observe a dramatically different behavior. For instance, considering $P=5$~bar, the isobar starts in the top right corner for a low chemical potential  (see the red symbols in Fig.~\ref{Fig3}). At this stage, the fluid is a supercritical vapor. As the chemical potential increases, the temperature and specific volume decrease, resulting in a decrease in entropy. This takes place until the isobar reaches the vapor-liquid coexistence curve (VLCC shown in cyan in Fig.~\ref{Fig3}) for a temperature $T=104.3 \pm 0.3$~K. Then, we observe a discontinuity in the isobar, marked by the red dashed line that shows the jump in entropy from the saturated vapor (cyan filled circle on the right side of the VLCC with $S=2.96 \pm 0.02$~kJ/kg) to the saturated liquid (cyan filled circle on the left side of the VLCC with $S=1.57 \pm 0.02$~kJ/kg). Then, the isobar continues on the left-hand (liquid) side of the phase diagram (see the red filled circle on the left of the VLCC in the bottom right corner of Fig.~\ref{Fig3}). Repeating this process for different pressures allows us to map out the whole VLCC (only a few isobars are shown for clarity in Fig.~\ref{Fig3}). We add that Fig.~\ref{Fig3} includes a comparison of the $(\mu,P,R)$ simulation results to the experimental data~\cite{Vargaftik} both for along the isobars and for the VLCC, which shows that $(\mu,P,R)$ can indeed predict reliably the onset and locus of the phase transition.

\begin{figure}
\begin{center}
\includegraphics*[width=10cm]{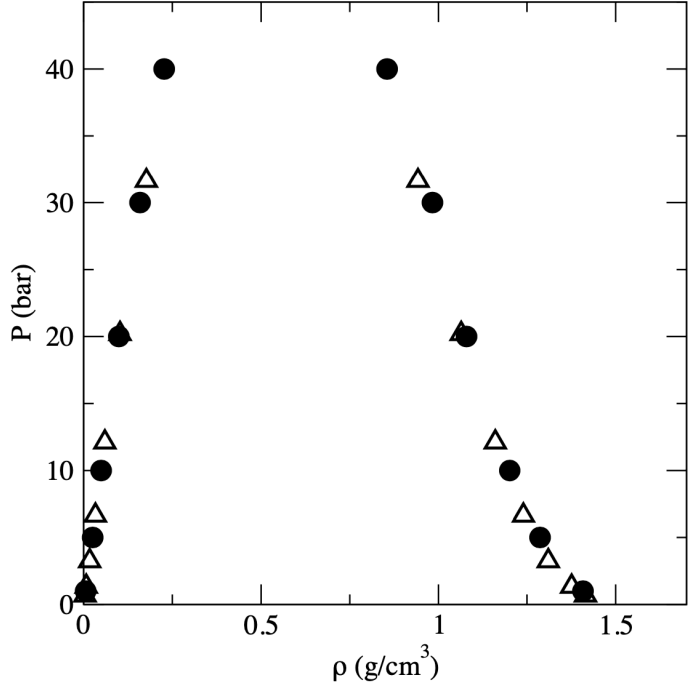}
\end{center}
\caption{Vapor-liquid equilibria for Argon: comparison between the pressure-density plots obtained using $(\mu,P,R)$ simulations (filled circles) and Expanded Wang-Landau simulations (open triangles).}
\label{Fig4}
\end{figure}

To assess further the accuracy of the results, we compare the $(\mu,P,R)$ simulation results to another set of simulation results obtained in previous work~\cite{PartI} with a flat-histogram simulation, the Expanded Wang-Landau (EWL) method. The two simulation methods rely on very different principles and, as such, provide a stringent test of the accuracy of the $(\mu,P,R)$ simulations. The EWL method is implemented in an isothermal ensemble, {\it i.e.} in the grand-canonical ensemble, and applied to on Argon, modeled with the Lennard-Jones and the same cutoff radius for the potential. Through the numerical determination of the partition function, the EWL method determines that phase coexistence is achieved when the two peaks in the number distribution $p(N)$ have equal probabilities. On the other hand, $(\mu,P,R)$ simulations determine that two phases of different densities $\rho_l$ and $\rho_v$ when they share the same input parameters for the chemical potential ($\mu_l=\mu_v$) and for the pressure ($P_l=P_v$) and the same output value for the temperature ($<T_l>=<T_v>$). The comparison is shown here for the vapor-liquid equilibria in the pressure-density plane, with Fig.~\ref{Fig4} establishing that the results from the two simulation methods are in very good agreement and thus that $(\mu,P,R)$ simulations provide an accurate account of the properties at coexistence.

\begin{figure}
\begin{center}
\includegraphics*[width=10cm]{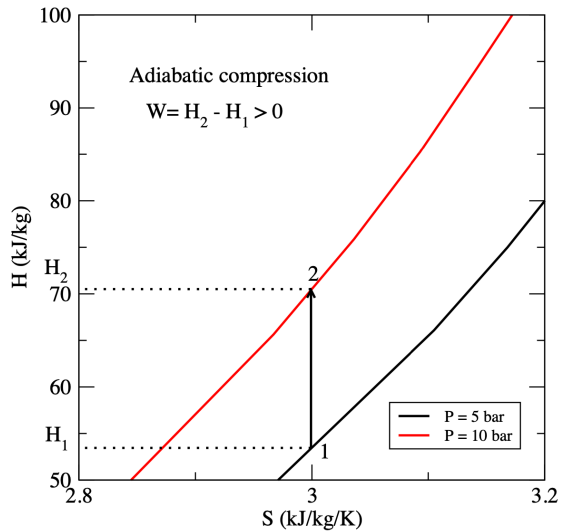}
\end{center}
\caption{$\mu$PR simulations of Argon: determination of the enthalpy change and of the work done on the gas during an adiabatic compression (isentropic process) from $5$~bar to $10$~bar.}
\label{Fig5}
\end{figure}

A significant advantage of simulations in an adiabatic ensemble, like the $(\mu,P,R)$ ensemble, is the connection that it provides with a number of engineering processes and devices, including nozzles, turbines and pumps. Indeed, as entropy is a readily available property during $(\mu,P,R)$ simulations, energy balances and enthalpy changes during isentropic processes can be evaluated using simulations in this ensemble. Since adiabatic, {\it i.e.} isentropic processes, are key steps in a number of idealized thermodynamic cycles, including the Carnot, Rankine, Diesel and Brayton cycles, such calculations are essential to the determination of the efficiency of these cycles. To illustrate this point, we plot in Fig.~\ref{Fig5} the results of $(\mu,P,R)$ simulations obtained for Argon along the two isobars $P=5$~bar and $P=10$~bar. Fig.~\ref{Fig5} shows the variation of the enthalpy as a function of entropy along the isobars, making the determination of the enthalpy change during, {\it e.g.}, an adiabatic compression straightforward. Considering an isentropic compression at $S=3$~kJ/kg from $5$~bar (State point $\#1$) to $10$~bar (State point $\#2$), we find an enthalpy change of $\Delta H=H_2-H_1=70.48-53.38=17.1\pm 0.2$~kJ/kg. The experimental data for Argon~\cite{Vargaftik} gives the following estimate for the enthalpy: $H^{exp}_1=52.7$~kJ/kg and $H^{exp}_2=69.8$~kJ/kg. This yields an enthalpy change of  $\Delta H=17.1$~kJ/kg, in excellent agreement between the experiment and the $(\mu,P,R)$ simulation results.

\begin{figure}
\begin{center}
\includegraphics*[width=10cm]{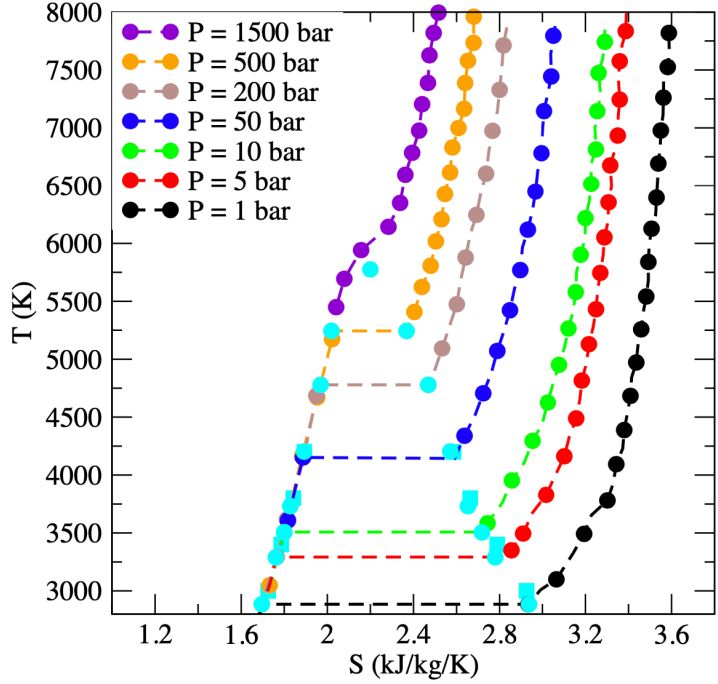}
\end{center}
\caption{Temperature-entropy plot for Copper. $(\mu,P,R)$ simulation results are shown as filled circles (colors vary according to the isobar they refer to, as indicated in the legend). Symbols in cyan outline the conditions for which vapor-liquid coexistence takes place according to the $(\mu,P,R)$ simulations (filled circles), while simulation results obtained in previous work~\cite{Leanna} are shown as squares. Only a few isobars are shown for clarity, while the $(\mu,P,R)$ results at coexistence are shown for all pressures considered here.}
\label{Fig6}
\end{figure}

We now perform $(\mu,P,R)$ simulations to determine the conditions for liquid-vapor equilibria for a metal and test the versatility of the approach. Having accurate and reliable simulation methods is especially important in the case of metals. It is indeed extremely difficult to carry out experiments close to the critical point for most metals, given the large temperatures and pressures involved, and there is generally a large uncertainty in their critical properties~\cite{schroer2014estimation,bhatt2006critical,apfelbaum2009predictions,apfelbaum2015similarity,Tsvetan1,metya2012molecular}. Here, we model Copper with the quantum corrected Sutton-Chen embedded atom model (qSC-EAM)~\cite{luo2003maximum} and follow the same procedure as for Argon to locate the vapor-liquid phase boundary. We thus follow the behavior of Copper along isobars ranging from $P=1$~bar to $P=1500$~bar. Starting from the top right corner of Fig.~\ref{Fig6}, we observe, depending on the isobar followed, two dramatically different behaviors. For isobars below $P=1500$~bars, we observe a steady decrease of entropy as $\mu$ increases along the isotherm, until the isobar reaches the vapor-liquid coexistence curve. At that point, we observe a discontinuity in the entropy, associated with the sudden change in density and thus entropy when the system goes from the vapor (high entropy branch of the binodal curve) to the liquid phase (low entropy branch of the binodal). In other words, Copper undergoes a first order phase transition, with discontinuities in entropy and molar volume at the transition. On the other hand, no such discontinuity is observed along the $P=1500$~bars, which indicates that Copper is a supercritical fluid under such conditions.

\begin{figure}
\begin{center}
\includegraphics*[width=10cm]{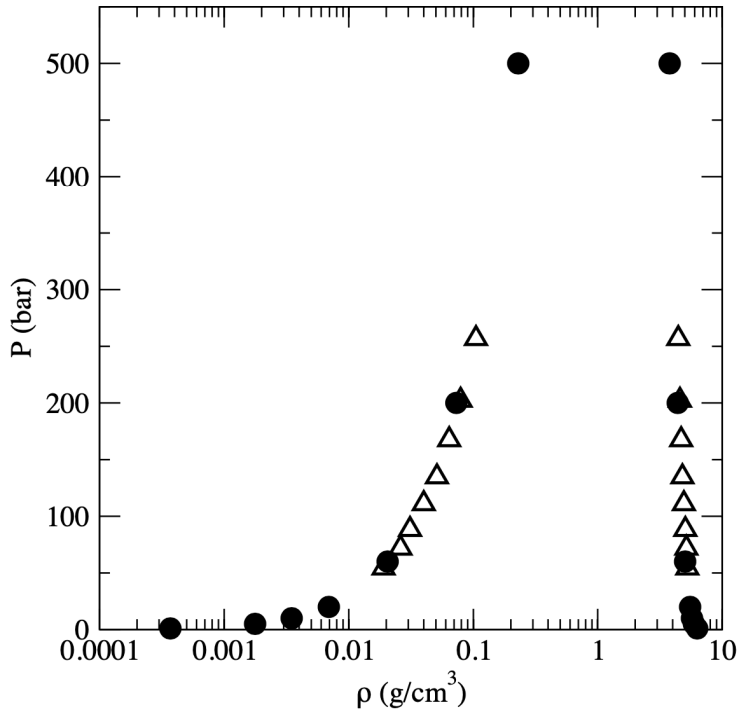}
\end{center}
\caption{(Vapor-liquid equilibria for Copper: comparison between the pressure-density plots obtained using $(\mu,P,R)$ simulations (filled circles) and Wang-Landau simulations~\cite{Tsvetan1} (open triangles).}
\label{Fig7}
\end{figure}

To explore further the relation between entropy and density, we present in Fig.~\ref{Fig7} the pressure-density plot obtained from the $(\mu,P,R)$ simulations. In particular, Fig.~\ref{Fig7} includes a comparison of the results from previous work using a flat-histogram method~\cite{Tsvetan1}. Both sets of results are in very good agreement, thereby confirming the accuracy and versatility of the $(\mu,P,R)$. The $(\mu,P,R)$ simulation data can then be used to determine an estimate of the critical temperature. Fitting the simulation results to the following functional form
\begin{equation}
\rho_l-\rho_v=B(T_c-T)^{0.325}
\end{equation}
in which $B$ and $T_c$ are two fitting parameters, provides a way to obtain the critical temperature (this functional form assumes a 3D-Ising critical exponent of 0.325). Applying this approach to the $(\mu,P,R)$ simulation results, we find a critical temperature $T_c=5775 \pm 30$~K, which is close to the prior estimate of $T_c=5696 \pm 50$~K obtained from flat-histogram simulations~\cite{Tsvetan1} for the qSC-EAM potential. This estimate is also within the range (between $5140$~K and $7696$~K) found in experiments~\cite{hess1998critical,martynyuk1976determination} and about $18$\% below the estimate ($7093$~K) found by extrapolating the experimental data for the density of the liquid phase at low temperature~\cite{apfelbaum2009predictions}. We also determine the critical pressure by fitting the ($\mu,P,R$) simulation results for the pressure-temperature relation to
\begin{equation}
\ln P = C - { D \over T+E}
\end{equation}
in which $C$, $D$ and $E$ are fitting parameters, and by calculating the value taken by pressure for a temperature equal to $T_c$. This yields a critical pressure $P_c=130\pm 15$~MPa, in reasonable agreement with findings from prior simulation work ($114 \pm 10$~MPa)~\cite{Tsvetan1}. It is also within the range of data found experimentally (between $42$~MPa and $583$~MPa)~\cite{hess1998critical,martynyuk1976determination}.

\begin{figure}
\begin{center}
\includegraphics*[width=7.75cm]{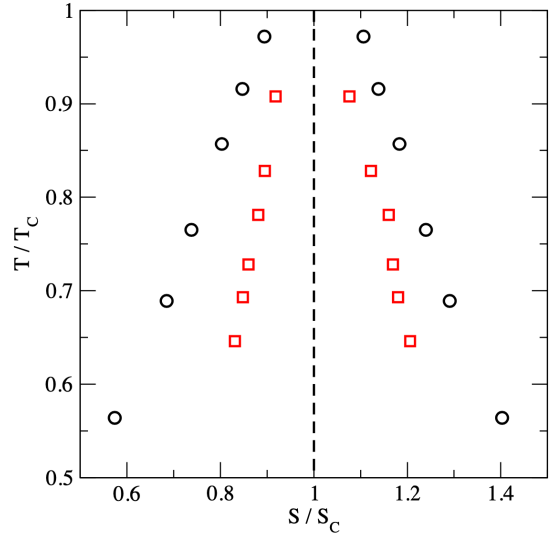}(a)
\includegraphics*[width=7.75cm]{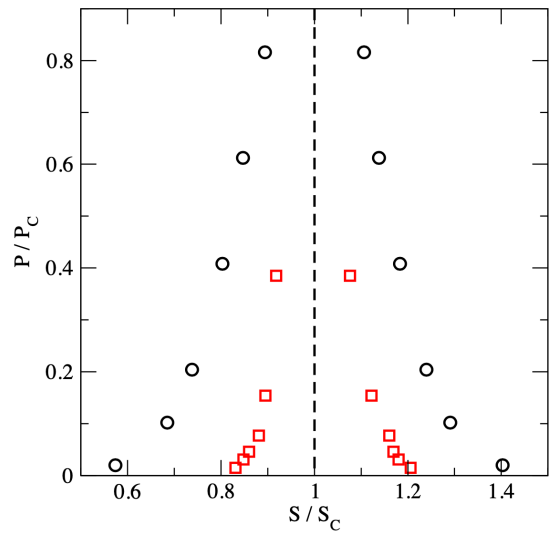}(b)
\end{center}
\caption{Scaled temperature-entropy (a) and pressure-entropy (b) plots for $Ar$ (black circles) and $Cu$ (red squares). The dashed vertical lines go through the critical point and are shown to highlight the symmetry of the plots.}
\label{Fig8}
\end{figure}

The estimates for $T_c$ and $P_c$ so obtained allow us to compare the behavior of the two systems through the scaling of their coexistence and critical properties. This is especially interesting since Argon is a system that obeys the law of corresponding states, while metals show departures from the generally observed behavior~\cite{bhatt2006critical,PhysRevLett.32.879}. This scaled plot will thus allow us to uncover, through the example of Argon, what the law of corresponding states entails from an entropic standpoint and to detect the onset of any anomalous behavior by comparing the behavior of Argon and Copper. For this purpose, we rescale the temperature and pressure by the critical parameters for each system. To scale the entropy, we estimate the critical entropy as follows. As shown in Figs.~\ref{Fig3} and~\ref{Fig6}, the coexistence point the closest to the critical point shows that the entropies of the two coexisting phases are within $0.5$~kJ/kg/K of each other. Thus, we take as the critical entropy the average of the entropies for the two coexisting phases for the highest temperature obtained in the simulations. This gives a critical entropy $S_c=2.29$~kJ/kg/K for Argon, which is very close to the experimental estimate of $2.261$~kJ/kg/K. Using the same method, we find a critical entropy of $S_c=2.2$~kJ/kg/K for Copper. This leads to the scaled temperature-entropy plot in Fig.~\ref{Fig8}(a) and to the scaled pressure-entropy plot in Fig.~\ref{Fig8}(b). For both systems, the results uncover a highly symmetric behavior exhibited by the two coexisting phases. This symmetry is in sharp contrast with the asymmetry observed in the pressure-density plots as in Figs.~\ref{Fig4}~and~\ref{Fig7}. We also observe that this symmetry applies to both Argon, a system that conforms to the law of corresponding states, and to Copper, which does not follows this law. This new insight in the phase transition process, when examined from an entropic standpoint, will be tested on an extensive range of systems in a forthcoming paper.

\section{Conclusions}
In this work, we leverage the central role played by entropy in adiabatic systems and show how simulations in the grand-isobaric adiabatic $(\mu,P,R)$ ensemble allow for a direct determination of entropy in bulk phases and along phase boundaries. Quantifying entropy during phase transitions and, more generally, order$\leftrightarrow$ disorder transitions is an especially timely issue as this concept has become essential to characterize~\cite{starr2003prediction,zha2016entropy}, follow~\cite{frenkel2014order} and even guide, through enhanced sampling~\cite{FS1,desgranges2017free,piaggi2017enhancing,desgranges2018crystal,doi:10.1080/08927022.2020.1761548}, self-organization and assembly processes in a wide range of systems.~\cite{tidor1994contribution,whitelam2015hierarchical,menzl2016effect,lee2018entropy,gobbo2018nucleation,martiniani2019quantifying}. In the $(\mu,P,R)$ ensemble, entropy can be directly obtained from the heat function $R$ and the application of the equipartition principle, which allows to detect first-order phase transitions, such as vapor-liquid equilibria, through its discontinuities and the onset of the supercritical fluid, through its switch to a continuous function. We perform Monte Carlo simulations in the $(\mu,P,R)$ ensemble on two examples, Argon and Copper, and assess the reliability of the method through comparison to experimental data and to the results from previous flat-histogram simulations. The results provide a picture of the phase transitions from an entropic standpoint, showing a remarkable symmetry with respect to the entropy of the coexisting phases as shown in scaled pressure-entropy and temperature-entropy plots. We find the results to hold both for Argon, a system that follows the law of corresponding states, and Copper, a system that exhibits departures from this law. Extensive testing of this finding on a wide range of systems will be the topic of future work.

\begin{acknowledgments}
Partial funding for this research was provided by NSF through award CHE-1955403. This work used the Extreme Science and Engineering Discovery Environment (XSEDE)~\cite{xsede}, which is supported by National Science Foundation grant number ACI-1548562, and used the Open Science Grid through allocation TG-CHE200063.
\end{acknowledgments}

\vspace{1 cm}

{\bf Data availability}
\vspace{0.5 cm}

The data that support the findings of this study are available from the corresponding author upon reasonable request.

\bibliography{AdiabS}

\begin{thebibliography}{73}
\expandafter\ifx\csname natexlab\endcsname\relax\def\natexlab#1{#1}\fi
\expandafter\ifx\csname bibnamefont\endcsname\relax
  \def\bibnamefont#1{#1}\fi
\expandafter\ifx\csname bibfnamefont\endcsname\relax
  \def\bibfnamefont#1{#1}\fi
\expandafter\ifx\csname citenamefont\endcsname\relax
  \def\citenamefont#1{#1}\fi
\expandafter\ifx\csname url\endcsname\relax
  \def\url#1{\texttt{#1}}\fi
\expandafter\ifx\csname urlprefix\endcsname\relax\def\urlprefix{URL }\fi
\providecommand{\bibinfo}[2]{#2}
\providecommand{\eprint}[2][]{\url{#2}}

\bibitem[{\citenamefont{Fowler and Guggenheim}(1939)}]{fowlerstatistical}
\bibinfo{author}{\bibfnamefont{R.}~\bibnamefont{Fowler}} \bibnamefont{and}
  \bibinfo{author}{\bibfnamefont{E.}~\bibnamefont{Guggenheim}},
  \emph{\bibinfo{title}{Statistical thermodynamics}}
  (\bibinfo{publisher}{Cambridge University Press, London},
  \bibinfo{year}{1939}).

\bibitem[{\citenamefont{Hill}(1986)}]{hill1986introduction}
\bibinfo{author}{\bibfnamefont{T.~L.} \bibnamefont{Hill}},
  \emph{\bibinfo{title}{An introduction to statistical thermodynamics}}
  (\bibinfo{publisher}{Dover Books, New York}, \bibinfo{year}{1986}).

\bibitem[{\citenamefont{Graben and Ray}(1993)}]{graben1993eight}
\bibinfo{author}{\bibfnamefont{H.}~\bibnamefont{Graben}} \bibnamefont{and}
  \bibinfo{author}{\bibfnamefont{J.~R.} \bibnamefont{Ray}},
  \bibinfo{journal}{Mol. Phys.} \textbf{\bibinfo{volume}{80}},
  \bibinfo{pages}{1183} (\bibinfo{year}{1993}).

\bibitem[{\citenamefont{Escobedo}(2006)}]{escobedo2006simulation}
\bibinfo{author}{\bibfnamefont{F.~A.} \bibnamefont{Escobedo}},
  \bibinfo{journal}{Phys. Rev. E} \textbf{\bibinfo{volume}{73}},
  \bibinfo{pages}{056701} (\bibinfo{year}{2006}).

\bibitem[{\citenamefont{McQuarrie}(1976)}]{McQuarrie}
\bibinfo{author}{\bibfnamefont{D.~A.} \bibnamefont{McQuarrie}},
  \emph{\bibinfo{title}{Statistical Mechanics}} (\bibinfo{publisher}{Harper \&
  Row, New York}, \bibinfo{year}{1976}).

\bibitem[{\citenamefont{Allen and Tildesley}(1987)}]{Allen}
\bibinfo{author}{\bibfnamefont{M.~P.} \bibnamefont{Allen}} \bibnamefont{and}
  \bibinfo{author}{\bibfnamefont{D.~J.} \bibnamefont{Tildesley}},
  \emph{\bibinfo{title}{Computer Simulation of Liquids}}
  (\bibinfo{publisher}{Clarendon, Oxford}, \bibinfo{year}{1987}).

\bibitem[{\citenamefont{Nos{\'e}}(1984)}]{nose1984molecular}
\bibinfo{author}{\bibfnamefont{S.}~\bibnamefont{Nos{\'e}}},
  \bibinfo{journal}{Mol. Phys.} \textbf{\bibinfo{volume}{52}},
  \bibinfo{pages}{255} (\bibinfo{year}{1984}).

\bibitem[{\citenamefont{Andersen}(1980)}]{andersen1980molecular}
\bibinfo{author}{\bibfnamefont{H.~C.} \bibnamefont{Andersen}},
  \bibinfo{journal}{J. Chem. Phys.} \textbf{\bibinfo{volume}{72}},
  \bibinfo{pages}{2384} (\bibinfo{year}{1980}).

\bibitem[{\citenamefont{{\c{C}}a{\u{g}}in and
  Pettitt}(1991)}]{ccaugin1991molecular}
\bibinfo{author}{\bibfnamefont{T.}~\bibnamefont{{\c{C}}a{\u{g}}in}}
  \bibnamefont{and} \bibinfo{author}{\bibfnamefont{B.~M.}
  \bibnamefont{Pettitt}}, \bibinfo{journal}{Mol. Phys.}
  \textbf{\bibinfo{volume}{72}}, \bibinfo{pages}{169} (\bibinfo{year}{1991}).

\bibitem[{\citenamefont{Martyna et~al.}(1994)\citenamefont{Martyna, Tobias, and
  Klein}}]{martyna1994constant}
\bibinfo{author}{\bibfnamefont{G.~J.} \bibnamefont{Martyna}},
  \bibinfo{author}{\bibfnamefont{D.~J.} \bibnamefont{Tobias}},
  \bibnamefont{and} \bibinfo{author}{\bibfnamefont{M.~L.} \bibnamefont{Klein}},
  \bibinfo{journal}{J. Chem. Phys.} \textbf{\bibinfo{volume}{101}},
  \bibinfo{pages}{4177} (\bibinfo{year}{1994}).

\bibitem[{\citenamefont{Martyna et~al.}(1996)\citenamefont{Martyna, Tuckerman,
  Tobias, and Klein}}]{martyna1996explicit}
\bibinfo{author}{\bibfnamefont{G.~J.} \bibnamefont{Martyna}},
  \bibinfo{author}{\bibfnamefont{M.~E.} \bibnamefont{Tuckerman}},
  \bibinfo{author}{\bibfnamefont{D.~J.} \bibnamefont{Tobias}},
  \bibnamefont{and} \bibinfo{author}{\bibfnamefont{M.~L.} \bibnamefont{Klein}},
  \bibinfo{journal}{Mol. Phys.} \textbf{\bibinfo{volume}{87}},
  \bibinfo{pages}{1117} (\bibinfo{year}{1996}).

\bibitem[{\citenamefont{Bond et~al.}(1999)\citenamefont{Bond, Leimkuhler, and
  Laird}}]{bond1999nose}
\bibinfo{author}{\bibfnamefont{S.~D.} \bibnamefont{Bond}},
  \bibinfo{author}{\bibfnamefont{B.~J.} \bibnamefont{Leimkuhler}},
  \bibnamefont{and} \bibinfo{author}{\bibfnamefont{B.~B.} \bibnamefont{Laird}},
  \bibinfo{journal}{J. Comput. Phys.} \textbf{\bibinfo{volume}{151}},
  \bibinfo{pages}{114} (\bibinfo{year}{1999}).

\bibitem[{\citenamefont{Bussi et~al.}(2007)\citenamefont{Bussi, Donadio, and
  Parrinello}}]{bussi2007canonical}
\bibinfo{author}{\bibfnamefont{G.}~\bibnamefont{Bussi}},
  \bibinfo{author}{\bibfnamefont{D.}~\bibnamefont{Donadio}}, \bibnamefont{and}
  \bibinfo{author}{\bibfnamefont{M.}~\bibnamefont{Parrinello}},
  \bibinfo{journal}{J. Chem. Phys.} \textbf{\bibinfo{volume}{126}},
  \bibinfo{pages}{014101} (\bibinfo{year}{2007}).

\bibitem[{\citenamefont{Tuckerman et~al.}(2006)\citenamefont{Tuckerman,
  Alejandre, L{\'o}pez-Rend{\'o}n, Jochim, and
  Martyna}}]{tuckerman2006liouville}
\bibinfo{author}{\bibfnamefont{M.~E.} \bibnamefont{Tuckerman}},
  \bibinfo{author}{\bibfnamefont{J.}~\bibnamefont{Alejandre}},
  \bibinfo{author}{\bibfnamefont{R.}~\bibnamefont{L{\'o}pez-Rend{\'o}n}},
  \bibinfo{author}{\bibfnamefont{A.~L.} \bibnamefont{Jochim}},
  \bibnamefont{and} \bibinfo{author}{\bibfnamefont{G.~J.}
  \bibnamefont{Martyna}}, \bibinfo{journal}{J. Phys. A: Math. Gen.}
  \textbf{\bibinfo{volume}{39}}, \bibinfo{pages}{5629} (\bibinfo{year}{2006}).

\bibitem[{\citenamefont{McDonald}(1972)}]{mcdonald1972npt}
\bibinfo{author}{\bibfnamefont{I.}~\bibnamefont{McDonald}},
  \bibinfo{journal}{Mol. Phys.} \textbf{\bibinfo{volume}{23}},
  \bibinfo{pages}{41} (\bibinfo{year}{1972}).

\bibitem[{\citenamefont{Jorgensen}(1982)}]{jorgensen1982convergence}
\bibinfo{author}{\bibfnamefont{W.~L.} \bibnamefont{Jorgensen}},
  \bibinfo{journal}{Chem. Phys. Lett.} \textbf{\bibinfo{volume}{92}},
  \bibinfo{pages}{405} (\bibinfo{year}{1982}).

\bibitem[{\citenamefont{Ray}(1991)}]{ray1991microcanonical}
\bibinfo{author}{\bibfnamefont{J.~R.} \bibnamefont{Ray}},
  \bibinfo{journal}{Phys. Rev. A} \textbf{\bibinfo{volume}{44}},
  \bibinfo{pages}{4061} (\bibinfo{year}{1991}).

\bibitem[{\citenamefont{Ray and Frel{\'e}choz}(1996)}]{ray1996microcanonical}
\bibinfo{author}{\bibfnamefont{J.~R.} \bibnamefont{Ray}} \bibnamefont{and}
  \bibinfo{author}{\bibfnamefont{C.}~\bibnamefont{Frel{\'e}choz}},
  \bibinfo{journal}{Phys. Rev. E} \textbf{\bibinfo{volume}{53}},
  \bibinfo{pages}{3402} (\bibinfo{year}{1996}).

\bibitem[{\citenamefont{Mezei}(1987)}]{mezei1987grand}
\bibinfo{author}{\bibfnamefont{M.}~\bibnamefont{Mezei}}, \bibinfo{journal}{Mol.
  Phys.} \textbf{\bibinfo{volume}{61}}, \bibinfo{pages}{565}
  (\bibinfo{year}{1987}).

\bibitem[{\citenamefont{Orkoulas and
  Panagiotopoulos}(1999)}]{orkoulas1999phase}
\bibinfo{author}{\bibfnamefont{G.}~\bibnamefont{Orkoulas}} \bibnamefont{and}
  \bibinfo{author}{\bibfnamefont{A.~Z.} \bibnamefont{Panagiotopoulos}},
  \bibinfo{journal}{J. Chem. Phys.} \textbf{\bibinfo{volume}{110}},
  \bibinfo{pages}{1581} (\bibinfo{year}{1999}).

\bibitem[{\citenamefont{Smit}(1995)}]{smit1995grand}
\bibinfo{author}{\bibfnamefont{B.}~\bibnamefont{Smit}}, \bibinfo{journal}{Mol.
  Phys.} \textbf{\bibinfo{volume}{85}}, \bibinfo{pages}{153}
  (\bibinfo{year}{1995}).

\bibitem[{\citenamefont{Jorgensen and Jenson}(1998)}]{jorgensen1998temperature}
\bibinfo{author}{\bibfnamefont{W.~L.} \bibnamefont{Jorgensen}}
  \bibnamefont{and} \bibinfo{author}{\bibfnamefont{C.}~\bibnamefont{Jenson}},
  \bibinfo{journal}{J. Comput. Chem.} \textbf{\bibinfo{volume}{19}},
  \bibinfo{pages}{1179} (\bibinfo{year}{1998}).

\bibitem[{\citenamefont{Errington and
  Panagiotopoulos}(1998)}]{errington1998phase}
\bibinfo{author}{\bibfnamefont{J.~R.} \bibnamefont{Errington}}
  \bibnamefont{and} \bibinfo{author}{\bibfnamefont{A.~Z.}
  \bibnamefont{Panagiotopoulos}}, \bibinfo{journal}{J. Chem. Phys.}
  \textbf{\bibinfo{volume}{109}}, \bibinfo{pages}{1093} (\bibinfo{year}{1998}).

\bibitem[{\citenamefont{Guggenheim}(1939)}]{guggenheim1939grand}
\bibinfo{author}{\bibfnamefont{E.}~\bibnamefont{Guggenheim}},
  \bibinfo{journal}{J. Chem. Phys.} \textbf{\bibinfo{volume}{7}},
  \bibinfo{pages}{103} (\bibinfo{year}{1939}).

\bibitem[{\citenamefont{Brown}(1958)}]{brown1958constant}
\bibinfo{author}{\bibfnamefont{W.~B.} \bibnamefont{Brown}},
  \bibinfo{journal}{Mol. Phys.} \textbf{\bibinfo{volume}{1}},
  \bibinfo{pages}{68} (\bibinfo{year}{1958}).

\bibitem[{\citenamefont{Haile and Graben}(1980)}]{haile1980isoenthalpic}
\bibinfo{author}{\bibfnamefont{J.}~\bibnamefont{Haile}} \bibnamefont{and}
  \bibinfo{author}{\bibfnamefont{H.}~\bibnamefont{Graben}},
  \bibinfo{journal}{Mol. Phys.} \textbf{\bibinfo{volume}{40}},
  \bibinfo{pages}{1433} (\bibinfo{year}{1980}).

\bibitem[{\citenamefont{Ray et~al.}(1981)\citenamefont{Ray, Graben, and
  Haile}}]{ray1981new}
\bibinfo{author}{\bibfnamefont{J.~R.} \bibnamefont{Ray}},
  \bibinfo{author}{\bibfnamefont{H.}~\bibnamefont{Graben}}, \bibnamefont{and}
  \bibinfo{author}{\bibfnamefont{J.}~\bibnamefont{Haile}}, \bibinfo{journal}{J.
  Chem. Phys.} \textbf{\bibinfo{volume}{75}}, \bibinfo{pages}{4077}
  (\bibinfo{year}{1981}).

\bibitem[{\citenamefont{Krist{\'o}f and Liszi}(1996)}]{kristof1996alternative}
\bibinfo{author}{\bibfnamefont{T.}~\bibnamefont{Krist{\'o}f}} \bibnamefont{and}
  \bibinfo{author}{\bibfnamefont{J.}~\bibnamefont{Liszi}},
  \bibinfo{journal}{Chem. Phys. Lett.} \textbf{\bibinfo{volume}{261}},
  \bibinfo{pages}{620} (\bibinfo{year}{1996}).

\bibitem[{\citenamefont{Graben and Ray}(1991)}]{graben1991unified}
\bibinfo{author}{\bibfnamefont{H.}~\bibnamefont{Graben}} \bibnamefont{and}
  \bibinfo{author}{\bibfnamefont{J.~R.} \bibnamefont{Ray}},
  \bibinfo{journal}{Phys. Rev. A} \textbf{\bibinfo{volume}{43}},
  \bibinfo{pages}{4100} (\bibinfo{year}{1991}).

\bibitem[{\citenamefont{Ray and Graben}(1990)}]{ray1990fourth}
\bibinfo{author}{\bibfnamefont{J.~R.} \bibnamefont{Ray}} \bibnamefont{and}
  \bibinfo{author}{\bibfnamefont{H.}~\bibnamefont{Graben}},
  \bibinfo{journal}{J. Chem. Phys.} \textbf{\bibinfo{volume}{93}},
  \bibinfo{pages}{4296} (\bibinfo{year}{1990}).

\bibitem[{\citenamefont{Ray and Wolf}(1993{\natexlab{a}})}]{ray1993monte}
\bibinfo{author}{\bibfnamefont{J.~R.} \bibnamefont{Ray}} \bibnamefont{and}
  \bibinfo{author}{\bibfnamefont{R.~J.} \bibnamefont{Wolf}},
  \bibinfo{journal}{J. Chem. Phys.} \textbf{\bibinfo{volume}{98}},
  \bibinfo{pages}{2263} (\bibinfo{year}{1993}{\natexlab{a}}).

\bibitem[{\citenamefont{Ray and Wolf}(1993{\natexlab{b}})}]{ray1993new}
\bibinfo{author}{\bibfnamefont{J.~R.} \bibnamefont{Ray}} \bibnamefont{and}
  \bibinfo{author}{\bibfnamefont{R.~J.} \bibnamefont{Wolf}}, in
  \emph{\bibinfo{booktitle}{Computer Simulation Studies in Condensed-Matter
  Physics VI. Springer Proceedings in Physics, vol. 76}}, edited by
  \bibinfo{editor}{\bibfnamefont{D.}~\bibnamefont{Landau}},
  \bibinfo{editor}{\bibfnamefont{K.}~\bibnamefont{Mon}}, \bibnamefont{and}
  \bibinfo{editor}{\bibfnamefont{H.}~\bibnamefont{Schuettler}}
  (\bibinfo{publisher}{Springer}, \bibinfo{address}{Berlin, Heidelberg},
  \bibinfo{year}{1993}{\natexlab{b}}).

\bibitem[{\citenamefont{Errington}(2003)}]{errington2003direct}
\bibinfo{author}{\bibfnamefont{J.~R.} \bibnamefont{Errington}},
  \bibinfo{journal}{J. Chem. Phys.} \textbf{\bibinfo{volume}{118}},
  \bibinfo{pages}{9915} (\bibinfo{year}{2003}).

\bibitem[{\citenamefont{Desgranges and Delhommelle}(2012)}]{PartI}
\bibinfo{author}{\bibfnamefont{C.}~\bibnamefont{Desgranges}} \bibnamefont{and}
  \bibinfo{author}{\bibfnamefont{J.}~\bibnamefont{Delhommelle}},
  \bibinfo{journal}{J. Chem. Phys.} \textbf{\bibinfo{volume}{136}},
  \bibinfo{pages}{184107} (\bibinfo{year}{2012}).

\bibitem[{\citenamefont{Vargaftik et~al.}(1996)\citenamefont{Vargaftik,
  Vinoradov, and Yargin}}]{Vargaftik}
\bibinfo{author}{\bibfnamefont{N.~B.} \bibnamefont{Vargaftik}},
  \bibinfo{author}{\bibfnamefont{Y.~K.} \bibnamefont{Vinoradov}},
  \bibnamefont{and} \bibinfo{author}{\bibfnamefont{V.~S.}
  \bibnamefont{Yargin}}, \emph{\bibinfo{title}{Handbook of Physical Properties
  of Liquids and Gases}} (\bibinfo{publisher}{Begell House, New York},
  \bibinfo{year}{1996}).

\bibitem[{\citenamefont{Finnis and Sinclair}(1984)}]{finnis1984simple}
\bibinfo{author}{\bibfnamefont{M.}~\bibnamefont{Finnis}} \bibnamefont{and}
  \bibinfo{author}{\bibfnamefont{J.}~\bibnamefont{Sinclair}},
  \bibinfo{journal}{Phil. Mag. A} \textbf{\bibinfo{volume}{50}},
  \bibinfo{pages}{45} (\bibinfo{year}{1984}).

\bibitem[{\citenamefont{Sutton and Chen}(1990)}]{sutton1990long}
\bibinfo{author}{\bibfnamefont{A.}~\bibnamefont{Sutton}} \bibnamefont{and}
  \bibinfo{author}{\bibfnamefont{J.}~\bibnamefont{Chen}},
  \bibinfo{journal}{Phil. Mag. Lett.} \textbf{\bibinfo{volume}{61}},
  \bibinfo{pages}{139} (\bibinfo{year}{1990}).

\bibitem[{\citenamefont{Mei et~al.}(1991)\citenamefont{Mei, Davenport, and
  Fernando}}]{mei1991analytic}
\bibinfo{author}{\bibfnamefont{J.}~\bibnamefont{Mei}},
  \bibinfo{author}{\bibfnamefont{J.}~\bibnamefont{Davenport}},
  \bibnamefont{and} \bibinfo{author}{\bibfnamefont{G.}~\bibnamefont{Fernando}},
  \bibinfo{journal}{Phys. Rev. B} \textbf{\bibinfo{volume}{43}},
  \bibinfo{pages}{4653} (\bibinfo{year}{1991}).

\bibitem[{\citenamefont{Daw and Baskes}(1983)}]{daw1983semiempirical}
\bibinfo{author}{\bibfnamefont{M.~S.} \bibnamefont{Daw}} \bibnamefont{and}
  \bibinfo{author}{\bibfnamefont{M.~I.} \bibnamefont{Baskes}},
  \bibinfo{journal}{Phys. Rev. Lett.} \textbf{\bibinfo{volume}{50}},
  \bibinfo{pages}{1285} (\bibinfo{year}{1983}).

\bibitem[{\citenamefont{Luo et~al.}(2003)\citenamefont{Luo, Ahrens,
  {\c{C}}a{\u{g}}{\i}n, Strachan, Goddard~III, and Swift}}]{luo2003maximum}
\bibinfo{author}{\bibfnamefont{S.-N.} \bibnamefont{Luo}},
  \bibinfo{author}{\bibfnamefont{T.~J.} \bibnamefont{Ahrens}},
  \bibinfo{author}{\bibfnamefont{T.}~\bibnamefont{{\c{C}}a{\u{g}}{\i}n}},
  \bibinfo{author}{\bibfnamefont{A.}~\bibnamefont{Strachan}},
  \bibinfo{author}{\bibfnamefont{W.~A.} \bibnamefont{Goddard~III}},
  \bibnamefont{and} \bibinfo{author}{\bibfnamefont{D.~C.} \bibnamefont{Swift}},
  \bibinfo{journal}{Phys. Rev. B} \textbf{\bibinfo{volume}{68}},
  \bibinfo{pages}{134206} (\bibinfo{year}{2003}).

\bibitem[{\citenamefont{Desgranges and
  Delhommelle}(2008)}]{desgranges2008rheology}
\bibinfo{author}{\bibfnamefont{C.}~\bibnamefont{Desgranges}} \bibnamefont{and}
  \bibinfo{author}{\bibfnamefont{J.}~\bibnamefont{Delhommelle}},
  \bibinfo{journal}{Phys. Rev. B} \textbf{\bibinfo{volume}{78}},
  \bibinfo{pages}{184202} (\bibinfo{year}{2008}).

\bibitem[{\citenamefont{Kart et~al.}(2005)\citenamefont{Kart, Tomak,
  Uludo{\u{g}}an, and {\c{C}}a{\u{g}}{\i}n}}]{kart2005thermodynamical}
\bibinfo{author}{\bibfnamefont{H.}~\bibnamefont{Kart}},
  \bibinfo{author}{\bibfnamefont{M.}~\bibnamefont{Tomak}},
  \bibinfo{author}{\bibfnamefont{M.}~\bibnamefont{Uludo{\u{g}}an}},
  \bibnamefont{and}
  \bibinfo{author}{\bibfnamefont{T.}~\bibnamefont{{\c{C}}a{\u{g}}{\i}n}},
  \bibinfo{journal}{Comput. Mater. Sci.} \textbf{\bibinfo{volume}{32}},
  \bibinfo{pages}{107} (\bibinfo{year}{2005}).

\bibitem[{\citenamefont{Xu et~al.}(2005)\citenamefont{Xu, Cagin, and
  Goddard~III}}]{xu2005assessment}
\bibinfo{author}{\bibfnamefont{P.}~\bibnamefont{Xu}},
  \bibinfo{author}{\bibfnamefont{T.}~\bibnamefont{Cagin}}, \bibnamefont{and}
  \bibinfo{author}{\bibfnamefont{W.~A.} \bibnamefont{Goddard~III}},
  \bibinfo{journal}{J. Chem. Phys.} \textbf{\bibinfo{volume}{123}},
  \bibinfo{pages}{104506} (\bibinfo{year}{2005}).

\bibitem[{\citenamefont{Desgranges and
  Delhommelle}(2014)}]{desgranges2014unraveling}
\bibinfo{author}{\bibfnamefont{C.}~\bibnamefont{Desgranges}} \bibnamefont{and}
  \bibinfo{author}{\bibfnamefont{J.}~\bibnamefont{Delhommelle}},
  \bibinfo{journal}{J. Am. Chem. Soc.} \textbf{\bibinfo{volume}{136}},
  \bibinfo{pages}{8145} (\bibinfo{year}{2014}).

\bibitem[{\citenamefont{Desgranges and
  Delhommelle}(2018{\natexlab{a}})}]{desgranges2018unusual}
\bibinfo{author}{\bibfnamefont{C.}~\bibnamefont{Desgranges}} \bibnamefont{and}
  \bibinfo{author}{\bibfnamefont{J.}~\bibnamefont{Delhommelle}},
  \bibinfo{journal}{Phys. Rev. Lett.} \textbf{\bibinfo{volume}{120}},
  \bibinfo{pages}{115701} (\bibinfo{year}{2018}{\natexlab{a}}).

\bibitem[{\citenamefont{Desgranges and Delhommelle}(2019)}]{desgranges2019can}
\bibinfo{author}{\bibfnamefont{C.}~\bibnamefont{Desgranges}} \bibnamefont{and}
  \bibinfo{author}{\bibfnamefont{J.}~\bibnamefont{Delhommelle}},
  \bibinfo{journal}{Phys. Rev. Lett.} \textbf{\bibinfo{volume}{123}},
  \bibinfo{pages}{195701} (\bibinfo{year}{2019}).

\bibitem[{\citenamefont{Gelb and Chakraborty}(2011)}]{gelb2011boiling}
\bibinfo{author}{\bibfnamefont{L.~D.} \bibnamefont{Gelb}} \bibnamefont{and}
  \bibinfo{author}{\bibfnamefont{S.~N.} \bibnamefont{Chakraborty}},
  \bibinfo{journal}{J. Chem. Phys.} \textbf{\bibinfo{volume}{135}},
  \bibinfo{pages}{224113} (\bibinfo{year}{2011}).

\bibitem[{\citenamefont{Aleksandrov et~al.}(2012)\citenamefont{Aleksandrov,
  Desgranges, and Delhommelle}}]{Tsvetan2}
\bibinfo{author}{\bibfnamefont{T.}~\bibnamefont{Aleksandrov}},
  \bibinfo{author}{\bibfnamefont{C.}~\bibnamefont{Desgranges}},
  \bibnamefont{and}
  \bibinfo{author}{\bibfnamefont{J.}~\bibnamefont{Delhommelle}},
  \bibinfo{journal}{Molec. Simul.} \textbf{\bibinfo{volume}{38}},
  \bibinfo{pages}{1265} (\bibinfo{year}{2012}).

\bibitem[{\citenamefont{Desgranges et~al.}(2016)\citenamefont{Desgranges,
  Widhalm, and Delhommelle}}]{Leanna}
\bibinfo{author}{\bibfnamefont{C.}~\bibnamefont{Desgranges}},
  \bibinfo{author}{\bibfnamefont{L.}~\bibnamefont{Widhalm}}, \bibnamefont{and}
  \bibinfo{author}{\bibfnamefont{J.}~\bibnamefont{Delhommelle}},
  \bibinfo{journal}{J. Phys. Chem. B} \textbf{\bibinfo{volume}{120}},
  \bibinfo{pages}{5255} (\bibinfo{year}{2016}).

\bibitem[{\citenamefont{Schr{\"o}er and
  Pottlacher}(2014)}]{schroer2014estimation}
\bibinfo{author}{\bibfnamefont{W.}~\bibnamefont{Schr{\"o}er}} \bibnamefont{and}
  \bibinfo{author}{\bibfnamefont{G.}~\bibnamefont{Pottlacher}},
  \bibinfo{journal}{High Temp.-High Press.} \textbf{\bibinfo{volume}{43}}
  (\bibinfo{year}{2014}).

\bibitem[{\citenamefont{Bhatt et~al.}(2006)\citenamefont{Bhatt, Jasper,
  Schultz, Siepmann, and Truhlar}}]{bhatt2006critical}
\bibinfo{author}{\bibfnamefont{D.}~\bibnamefont{Bhatt}},
  \bibinfo{author}{\bibfnamefont{A.~W.} \bibnamefont{Jasper}},
  \bibinfo{author}{\bibfnamefont{N.~E.} \bibnamefont{Schultz}},
  \bibinfo{author}{\bibfnamefont{J.~I.} \bibnamefont{Siepmann}},
  \bibnamefont{and} \bibinfo{author}{\bibfnamefont{D.~G.}
  \bibnamefont{Truhlar}}, \bibinfo{journal}{J. Am. Chem. Soc.}
  \textbf{\bibinfo{volume}{128}}, \bibinfo{pages}{4224} (\bibinfo{year}{2006}).

\bibitem[{\citenamefont{Apfelbaum and
  VorobÕev}(2009)}]{apfelbaum2009predictions}
\bibinfo{author}{\bibfnamefont{E.}~\bibnamefont{Apfelbaum}} \bibnamefont{and}
  \bibinfo{author}{\bibfnamefont{V.}~\bibnamefont{VorobÕev}},
  \bibinfo{journal}{Chem. Phys. Lett.} \textbf{\bibinfo{volume}{467}},
  \bibinfo{pages}{318} (\bibinfo{year}{2009}).

\bibitem[{\citenamefont{Apfelbaum and
  VorobÕev}(2015)}]{apfelbaum2015similarity}
\bibinfo{author}{\bibfnamefont{E.}~\bibnamefont{Apfelbaum}} \bibnamefont{and}
  \bibinfo{author}{\bibfnamefont{V.}~\bibnamefont{VorobÕev}},
  \bibinfo{journal}{J. Phys. Chem. B} \textbf{\bibinfo{volume}{119}},
  \bibinfo{pages}{8419} (\bibinfo{year}{2015}).

\bibitem[{\citenamefont{Aleksandrov et~al.}(2010)\citenamefont{Aleksandrov,
  Desgranges, and Delhommelle}}]{Tsvetan1}
\bibinfo{author}{\bibfnamefont{T.}~\bibnamefont{Aleksandrov}},
  \bibinfo{author}{\bibfnamefont{C.}~\bibnamefont{Desgranges}},
  \bibnamefont{and}
  \bibinfo{author}{\bibfnamefont{J.}~\bibnamefont{Delhommelle}},
  \bibinfo{journal}{Fluid Phase Equil.} \textbf{\bibinfo{volume}{287}},
  \bibinfo{pages}{79} (\bibinfo{year}{2010}).

\bibitem[{\citenamefont{Metya et~al.}(2012)\citenamefont{Metya, Hens, and
  Singh}}]{metya2012molecular}
\bibinfo{author}{\bibfnamefont{A.~K.} \bibnamefont{Metya}},
  \bibinfo{author}{\bibfnamefont{A.}~\bibnamefont{Hens}}, \bibnamefont{and}
  \bibinfo{author}{\bibfnamefont{J.~K.} \bibnamefont{Singh}},
  \bibinfo{journal}{Fluid Phase Equil.} \textbf{\bibinfo{volume}{313}},
  \bibinfo{pages}{16} (\bibinfo{year}{2012}).

\bibitem[{\citenamefont{Hess}(1998)}]{hess1998critical}
\bibinfo{author}{\bibfnamefont{H.}~\bibnamefont{Hess}}, \bibinfo{journal}{Z.
  Metallkd.} \textbf{\bibinfo{volume}{89}}, \bibinfo{pages}{388}
  (\bibinfo{year}{1998}).

\bibitem[{\citenamefont{Martynyuk and
  Pantelejchuk}(1976)}]{martynyuk1976determination}
\bibinfo{author}{\bibfnamefont{M.}~\bibnamefont{Martynyuk}} \bibnamefont{and}
  \bibinfo{author}{\bibfnamefont{O.}~\bibnamefont{Pantelejchuk}},
  \bibinfo{journal}{High Temp.-High Press.} \textbf{\bibinfo{volume}{14}},
  \bibinfo{pages}{1201} (\bibinfo{year}{1976}).

\bibitem[{\citenamefont{Weiner et~al.}(1974)\citenamefont{Weiner, Langley, and
  Ford}}]{PhysRevLett.32.879}
\bibinfo{author}{\bibfnamefont{J.}~\bibnamefont{Weiner}},
  \bibinfo{author}{\bibfnamefont{K.~H.} \bibnamefont{Langley}},
  \bibnamefont{and} \bibinfo{author}{\bibfnamefont{N.~C.} \bibnamefont{Ford}},
  \bibinfo{journal}{Phys. Rev. Lett.} \textbf{\bibinfo{volume}{32}},
  \bibinfo{pages}{879} (\bibinfo{year}{1974}).

\bibitem[{\citenamefont{Starr et~al.}(2003)\citenamefont{Starr, Angell, and
  Stanley}}]{starr2003prediction}
\bibinfo{author}{\bibfnamefont{F.~W.} \bibnamefont{Starr}},
  \bibinfo{author}{\bibfnamefont{C.~A.} \bibnamefont{Angell}},
  \bibnamefont{and} \bibinfo{author}{\bibfnamefont{H.~E.}
  \bibnamefont{Stanley}}, \bibinfo{journal}{Physica A}
  \textbf{\bibinfo{volume}{323}}, \bibinfo{pages}{51} (\bibinfo{year}{2003}).

\bibitem[{\citenamefont{Zha et~al.}(2016)\citenamefont{Zha, Zhang, Li, and
  Hu}}]{zha2016entropy}
\bibinfo{author}{\bibfnamefont{L.}~\bibnamefont{Zha}},
  \bibinfo{author}{\bibfnamefont{M.}~\bibnamefont{Zhang}},
  \bibinfo{author}{\bibfnamefont{L.}~\bibnamefont{Li}}, \bibnamefont{and}
  \bibinfo{author}{\bibfnamefont{W.}~\bibnamefont{Hu}}, \bibinfo{journal}{J.
  Phys. Chem. B} \textbf{\bibinfo{volume}{120}}, \bibinfo{pages}{12988}
  (\bibinfo{year}{2016}).

\bibitem[{\citenamefont{Frenkel}(2014)}]{frenkel2014order}
\bibinfo{author}{\bibfnamefont{D.}~\bibnamefont{Frenkel}},
  \bibinfo{journal}{Nat. Mater.} \textbf{\bibinfo{volume}{14}},
  \bibinfo{pages}{9} (\bibinfo{year}{2014}).

\bibitem[{\citenamefont{Desgranges and Delhommelle}(2016)}]{FS1}
\bibinfo{author}{\bibfnamefont{C.}~\bibnamefont{Desgranges}} \bibnamefont{and}
  \bibinfo{author}{\bibfnamefont{J.}~\bibnamefont{Delhommelle}},
  \bibinfo{journal}{J. Chem. Phys.} \textbf{\bibinfo{volume}{145}},
  \bibinfo{pages}{204112} (\bibinfo{year}{2016}).

\bibitem[{\citenamefont{Desgranges and Delhommelle}(2017)}]{desgranges2017free}
\bibinfo{author}{\bibfnamefont{C.}~\bibnamefont{Desgranges}} \bibnamefont{and}
  \bibinfo{author}{\bibfnamefont{J.}~\bibnamefont{Delhommelle}},
  \bibinfo{journal}{J. Chem. Phys.} \textbf{\bibinfo{volume}{146}},
  \bibinfo{pages}{184104} (\bibinfo{year}{2017}).

\bibitem[{\citenamefont{Piaggi et~al.}(2017)\citenamefont{Piaggi, Valsson, and
  Parrinello}}]{piaggi2017enhancing}
\bibinfo{author}{\bibfnamefont{P.~M.} \bibnamefont{Piaggi}},
  \bibinfo{author}{\bibfnamefont{O.}~\bibnamefont{Valsson}}, \bibnamefont{and}
  \bibinfo{author}{\bibfnamefont{M.}~\bibnamefont{Parrinello}},
  \bibinfo{journal}{Phys. Rev. Lett.} \textbf{\bibinfo{volume}{119}},
  \bibinfo{pages}{015701} (\bibinfo{year}{2017}).

\bibitem[{\citenamefont{Desgranges and
  Delhommelle}(2018{\natexlab{b}})}]{desgranges2018crystal}
\bibinfo{author}{\bibfnamefont{C.}~\bibnamefont{Desgranges}} \bibnamefont{and}
  \bibinfo{author}{\bibfnamefont{J.}~\bibnamefont{Delhommelle}},
  \bibinfo{journal}{Phys. Rev. E} \textbf{\bibinfo{volume}{98}},
  \bibinfo{pages}{063307} (\bibinfo{year}{2018}{\natexlab{b}}).

\bibitem[{\citenamefont{Tsai and
  Tiwary}(2020)}]{doi:10.1080/08927022.2020.1761548}
\bibinfo{author}{\bibfnamefont{S.-T.} \bibnamefont{Tsai}} \bibnamefont{and}
  \bibinfo{author}{\bibfnamefont{P.}~\bibnamefont{Tiwary}},
  \bibinfo{journal}{Molec. Simul.} \textbf{\bibinfo{volume}{DOI:
  10.1080/08927022.2020.1761548}} (\bibinfo{year}{2020}),
  \eprint{https://doi.org/10.1080/08927022.2020.1761548}.

\bibitem[{\citenamefont{Tidor and Karplus}(1994)}]{tidor1994contribution}
\bibinfo{author}{\bibfnamefont{B.}~\bibnamefont{Tidor}} \bibnamefont{and}
  \bibinfo{author}{\bibfnamefont{M.}~\bibnamefont{Karplus}},
  \bibinfo{journal}{J. Mol. Biol.} \textbf{\bibinfo{volume}{238}},
  \bibinfo{pages}{405} (\bibinfo{year}{1994}).

\bibitem[{\citenamefont{Whitelam}(2015)}]{whitelam2015hierarchical}
\bibinfo{author}{\bibfnamefont{S.}~\bibnamefont{Whitelam}},
  \bibinfo{journal}{Soft Matter} \textbf{\bibinfo{volume}{11}},
  \bibinfo{pages}{8225} (\bibinfo{year}{2015}).

\bibitem[{\citenamefont{Menzl and Dellago}(2016)}]{menzl2016effect}
\bibinfo{author}{\bibfnamefont{G.}~\bibnamefont{Menzl}} \bibnamefont{and}
  \bibinfo{author}{\bibfnamefont{C.}~\bibnamefont{Dellago}},
  \bibinfo{journal}{J. Chem. Phys.} \textbf{\bibinfo{volume}{145}},
  \bibinfo{pages}{211918} (\bibinfo{year}{2016}).

\bibitem[{\citenamefont{Lee et~al.}(2018)\citenamefont{Lee, Engel, and
  Glotzer}}]{lee2018entropy}
\bibinfo{author}{\bibfnamefont{S.}~\bibnamefont{Lee}},
  \bibinfo{author}{\bibfnamefont{M.}~\bibnamefont{Engel}}, \bibnamefont{and}
  \bibinfo{author}{\bibfnamefont{S.}~\bibnamefont{Glotzer}},
  \bibinfo{journal}{Bull. Am. Phys. Soc.} \textbf{\bibinfo{volume}{K57}},
  \bibinfo{pages}{00005} (\bibinfo{year}{2018}).

\bibitem[{\citenamefont{Gobbo et~al.}(2018)\citenamefont{Gobbo, Bellucci,
  Tribello, Ciccotti, and Trout}}]{gobbo2018nucleation}
\bibinfo{author}{\bibfnamefont{G.}~\bibnamefont{Gobbo}},
  \bibinfo{author}{\bibfnamefont{M.~A.} \bibnamefont{Bellucci}},
  \bibinfo{author}{\bibfnamefont{G.~A.} \bibnamefont{Tribello}},
  \bibinfo{author}{\bibfnamefont{G.}~\bibnamefont{Ciccotti}}, \bibnamefont{and}
  \bibinfo{author}{\bibfnamefont{B.~L.} \bibnamefont{Trout}},
  \bibinfo{journal}{J. Chem. Theory Comput.} \textbf{\bibinfo{volume}{14}},
  \bibinfo{pages}{959} (\bibinfo{year}{2018}).

\bibitem[{\citenamefont{Martiniani et~al.}(2019)\citenamefont{Martiniani,
  Chaikin, and Levine}}]{martiniani2019quantifying}
\bibinfo{author}{\bibfnamefont{S.}~\bibnamefont{Martiniani}},
  \bibinfo{author}{\bibfnamefont{P.~M.} \bibnamefont{Chaikin}},
  \bibnamefont{and} \bibinfo{author}{\bibfnamefont{D.}~\bibnamefont{Levine}},
  \bibinfo{journal}{Phys. Rev. X} \textbf{\bibinfo{volume}{9}},
  \bibinfo{pages}{011031} (\bibinfo{year}{2019}).

\bibitem[{\citenamefont{Towns et~al.}(2014)\citenamefont{Towns, Cockerill,
  Dahan, Foster, Gaither, Grimshaw, Hazlewood, Lathrop, Lifka, Peterson
  et~al.}}]{xsede}
\bibinfo{author}{\bibfnamefont{J.}~\bibnamefont{Towns}},
  \bibinfo{author}{\bibfnamefont{T.}~\bibnamefont{Cockerill}},
  \bibinfo{author}{\bibfnamefont{M.}~\bibnamefont{Dahan}},
  \bibinfo{author}{\bibfnamefont{I.}~\bibnamefont{Foster}},
  \bibinfo{author}{\bibfnamefont{K.}~\bibnamefont{Gaither}},
  \bibinfo{author}{\bibfnamefont{A.}~\bibnamefont{Grimshaw}},
  \bibinfo{author}{\bibfnamefont{V.}~\bibnamefont{Hazlewood}},
  \bibinfo{author}{\bibfnamefont{S.}~\bibnamefont{Lathrop}},
  \bibinfo{author}{\bibfnamefont{D.}~\bibnamefont{Lifka}},
  \bibinfo{author}{\bibfnamefont{G.~D.} \bibnamefont{Peterson}},
  \bibnamefont{et~al.}, \bibinfo{journal}{Computing in Science \& Engineering}
  \textbf{\bibinfo{volume}{16}}, \bibinfo{pages}{62} (\bibinfo{year}{2014}),
  ISSN \bibinfo{issn}{1521-9615},
  \urlprefix\url{doi.ieeecomputersociety.org/10.1109/MCSE.2014.80}.

\end{thebibliography}

\end{document}